# Navigating Unmeasured Confounding in Quantitative Sociology: A Sensitivity Framework

Cheng Lin*[1], Jose M. Peña[2], Adel Daoud[1,3]

1. Institute for Analytical Sociology (IAS), Linköping University, Linköping, Sweden
2. Department of Computer and Information Science (IDA), Linköping University, Linköping, Sweden
3. The Division of Data Science and Artificial Intelligence of the Department of Computer Science and Engineering, Chalmers University of Technology, Gothenburg, Sweden

*Corresponding author

**Acknowledgement:** We thank Sourabh Balgi for his efforts to contribute to reviewing sensitivity analysis methods. We also thank Carlos Cinelli, Christopher Winship, Connor T. Jerzak, Geoffrey Wodtke, and Ethan Fosse for helpful feedback and comments. In this article, we also utilized ChatGPT-4o and o1 to assist with language editing, as well as to test and modify R codes for examples and to test and modify LaTeX for creating figures. This work was supported by generous grants from the Swedish Research Council (VR); dnr. 2013-07681 and 2019-00245.

## Abstract

Unmeasured confounding remains a critical challenge in causal inference for the social sciences. This paper proposes a sensitivity analysis framework to systematically evaluate how unmeasured confounders influence statistical inference in sociology. Given these sensitivity analysis methods, we introduce a five-step workflow that integrates sensitivity analysis into research design rather than treating it as a post-hoc robustness check. Using the Blau and Duncan (1967) study as an empirical example, we demonstrate how different sensitivity methods provide complementary insights. By extending existing frameworks, we show how sensitivity analysis enhances causal transparency, offering a practical tool for assessing uncertainty in observational research. Our approach contributes to a more rigorous application of causal inference in sociology, bridging gaps between theory, identification strategies, and statistical modeling.

## 1. Introduction

Causal inference is one of the methodological foundations of applied sociology(Aalen et al. 2014; Gangl 2010; Hedstrom and Swedberg 1998; Hedström and Ylikoski 2010; Morgan and Winship 2015). Nevertheless, sociologists often refrain from making strong claims, particularly causal claims. One of the main reasons is the difficulties of addressing unobserved confounding (or omitted variable bias). In observational research, large parts of the relevant context remain unobserved, making it difficult to isolate a specific causal path of interest. As Lundberg, Johnson, and Stewart (2021) note, clearly defining one's causal estimand is crucial, but even a well-defined estimand can be undermined by unmeasured confounders. Unmeasured confounding violates the "unconfoundedness" assumption that all variables affecting both treatment and outcome have been accounted for. (Blum, Tan, and Ioannidis 2020; Frank et al. 2013; MacLehose et al. 2021; Manski 1990; Rosenbaum 2002a). In other words, unmeasured confounding represents an inherent uncertainty in social research– a reflection of the complex reality where we inevitably have incomplete knowledge of all causally relevant factors —whether in terms of theoretical mechanisms, model assumptions, or data constraints (Daoud and Dubhashi 2021; Deaton 2010; Morgan and Winship 2015). Importantly, investigating unobserved confounding is not about "detecting something bad" but about learning how hidden factors might be influencing our results, and thereby exercising epistemic humility about causal conclusions.

To systematically explore the role of unmeasured confounding in linking theory and statistical evidence, we decompose the unmeasured confounding question into three different components aligned to Lundberg et al.'s (2021) framework (1) how unmeasured confounding could affect the understanding of theoretical estimands (**confounding-estimand sensitivity**). (2) how identification strategies rely on assumptions about unmeasured confounders (**identification-assumption sensitivity**). (3) how statistical model specifications interact with the specification of unmeasured confounding (**specification sensitivity**). Rather than treating sensitivity analysis as a one-time post-hoc approach, we argue that it should be embedded throughout the causal inference process. Researchers should assess the impact of unmeasured confounding at each stage– from defining estimands, to choosing identification strategies, to specifying statistical models– thereby shifting the focus from eliminating confounding (often unrealistic in observational studies) to assessing the extent to which it could affect conclusions.

Under this sensitivity framework, building on a systematic review of sensitivity analysis techniques up to April 2024, we develop a heuristic framework that guides researchers in selecting the most appropriate method based on their estimand, data structure, and the scale of unmeasured confounding.

Sensitivity analysis is a technique that employs a series of tools to evaluate the resilience of causal relationships against unmeasured confounding, essentially conducting a thought experiment of the following form. Sensitivity analysis embodies a form of epistemic humility: through thought experiments and counterfactual reasoning, it systematically assesses how unmeasured factors might influence causal effects. It introduces hypothetical unobserved confounding into the data and gauges how swiftly the causal quantity of interest diminishes (Greenland 1996; Imbens and Rubin 2015; Rosenbaum 2002b). This technique echo debates and developments from the 1960s spearheaded by, among others, sociologists like Otis Duncan, Robert Hauser, and Arthur Goldberger. The latter two wrote in 1971, The Treatment of Unobservable Variables in Path Analysis, showing a method for sensitivity analysis in structural equation models (Hauser and Goldberger 1971). The employment of structural equation models during this period marked a significant methodological advancement in the articulation and testing of sociological theories, yet the adoption and application of these insights have been sporadic at best (Asparouhov, Hamaker, and Muthén 2018; Bielby and Hauser 1977; Harding 2003; Pearl 1998; Russo, Wunsch, and Mouchart 2019; Tarka 2018). High-quality applications already exist, yet many researchers who could benefit from these methods have not widely adopted them (Cinelli and Hazlett 2020; Imai, Keele, and Tingley 2010; Morgan and Winship 2015; VanderWeele 2010; Van Der Weele and Ding 2017; Wodtke, Elwert, and Harding 2016; Zhou and Yamamoto 2022).

By integrating the sensitivity framework with heuristic decision-making workflow of sensitivity analysis techniques, we offer a practical tool that enables researchers to: (1) Identify how different types of sensitivity affect their studies 2) Explore decision-making workflow in sensitivity analysis methods based on the scale of variables and the causal estimand of interest. (3) Interpret the implications of sensitivity analysis results in light of theoretical assumptions.

Therefore, the main contribution of this article includes two main parts: (1) We propose a sensitivity framework to explore the roles of unmeasured confounding in statistical inference. This expands sensitivity analysis beyond unmeasured confounding to recognize broader limitations in causal thinking. (2) We design a practical workflow that aligns different decision-making in sensitivity methods. Unlike previous approaches that treated sensitivity analysis as a post-hoc robustness check, our framework integrates it into the research design phase, making it a proactive tool for causal assessment.

## 2. The Missing Component of Estimand: Sensitivity Analysis and Unobserved Confounding

### 2.1 The usage of Sensitivity Analysis Terminology

Although sensitivity analysis methods have existed for decades, they have not gained the prominence one might expect in applied quantitative sociology. Adoption has been slow, in part because developments have been scattered across technical fields – biostatistics (VanderWeele and Ding 2017), statistics (Cinelli et al. 2019; Rosenbaum 2023), and computer science (Balgi, Peña, and Daoud 2024)—often in technical language that is difficult for applied sociologists to digest. As a result, few sociologists incorporate these tools routinely. (Cinelli and Hazlett 2020, 2022; Rosenbaum 2023; VanderWeele and Ding 2017).

We reviewed the prevalence of usage of the term sensitivity analysis and its relevant terms in the titles, abstracts, and keywords of papers in four Web of Science categories: "political science", "Economics", "sociology" and "Public, Environmental & Occupational Health" from 2002 to 2022. Figure 1 shows the usage of "sensitivity analysis" terms in different disciplines. There are increasing trends in the "Public, Environmental & Occupational Health" and "Economics" categories. However, "political science" and "sociology" are lagging. Please check Appendix A to find out how we search for this information. This broad search across disciplines provides a comparison of the uptake of sensitivity analysis, even only considering sensitivity analysis as a post-hoc methodological approach.

[Insert Figure 1 here]

Why might this be? One reason is that sociological research inherently involves multiple layers of uncertainty– from theoretical mechanisms and model assumptions to data limitations and external validity. Achieving complete certainty in observational studies is nearly impossible, and sociologists are keenly aware of this. Lundberg et al. (2021) provided a structured framework for defining inference goals in causal analysis. However, their framework implicitly assumes that once the estimand is defined, the uncertainty introduced by unmeasured confounding is second-order. Figure 2 extends their framework by illustrating where unmeasured confounding can creep in and influence key decisions in the research process.

[Insert Figure 2 here]

We categorize the impact of unmeasured confounding into three key dimensions of sensitivity analysis: (1) how unmeasured confounding influences the interpretation of theoretical estimands (confounding-estimand sensitivity), (2) how identification strategies depend on assumptions about unmeasured confounding (identification-assumption sensitivity), and (3) how statistical model choices interact with unmeasured confounding (specification sensitivity).

## 2.2 Confounding-Estimand Sensitivity

In sociological research, scholars often seek to identify a single causal path from X to Y. However, such an approach is at risk of capturing only a fraction of the full story in observational studies. This limitation renders causal explanations inherently incomplete, as identifying all latent causal paths is nearly impossible—even within the most "comprehensive" sociological theories. Many theoretical frameworks may overlook certain causally relevant processes or pathways, leaving some factors (U) unobserved. In this case, unmeasured confounders open alternative pathways—backdoor paths—that may be causal in other estimands but are non-causal with respect to the treatment–outcome relationship of interest. *These paths threaten the validity of our causal estimates and influence how researchers define their estimand and target population, a concern we term, which we term confounding-estimand sensitivity* (Cinelli and Hazlett 2020; Rosenbaum 2002a).

Multiple conditions are likely to interact to shape social outcomes (Morgan and Winship 2015; Ragin 1987), which makes it difficult to isolate a specific path in observational studies, particularly in variable-based research settings. Furthermore, the complexity of the relationship between theory and reality arises from the multilayered nature of causal mechanisms, which implies that researchers may not fully capture the dynamics, spanning from micro to macro levels (Hedstrom and Swedberg 1998). Even when adopting the "sometimes true" approach—where theories are understood as context-dependent (Hedström 2021)—researchers must still account for the influence of unmeasured confounders, because even context-dependent explanations require clear causal isolation. Thus, across these theoretical traditions, the presence of unmeasured confounding is not just about a set of unmeasured variables, but also likely to be the failure of capturing a path (either causal or non-causal), which may invalidate estimand. Rather than assuming away all confounding via design or controls, confounding–estimand sensitivity analysis invites investigators to theorize about $U$– what it could be, how it relates to $X$ and $Y$– and to quantify how much $U$ would need to matter to threaten the focal causal link.

Lundberg et al. (2021) provided a structured approach to defining estimands and linking theory to statistical evidence. However, their framework implicitly assumes that the target population and unit-specific quantity are not systematically biased by unmeasured confounders or at least does not explicitly discuss how unmeasured confounding could impact them. We extend their framework by demonstrating how unmeasured confounders can distort these definitions, leading to challenges in both generalizability (scope bias) and internal validity (subgroup heterogeneity).

When we suspect the presence of an unobserved confounder $U$ that affects both the treatment and the outcome, how "unit-specific quantity" bridges the link between theoretical goals and estimands becomes nuanced (Imbens and Rubin 2015; Lundberg et al. 2021). First, unmeasured confounders can distort the definition of the target population by systematically influencing both treatment assignment and selection into the study sample (Lesko et al. 2017). When an unobserved variable affects both who receives treatment and who is included in the analysis, the estimated causal effect may no longer generalize to the originally intended population, where a specific mechanism or theory may only work for a specific population rather than the originally intended population. For example, in studies of education and earnings, if unobserved family background influences both access to higher education and the likelihood of responding to surveys, then the estimated treatment effect of education on income may only apply to privileged individuals, rather than the general population of all potential collegegoers.

Second, unmeasured confounders can introduce bias in at least two ways when estimating subgroup effects. First, different subgroups may face different types of unobserved confounders, for instance, risk preferences in one group and family constraints in another—suggesting that the underlying data-generating mechanisms differ. Second, even when the same unobserved confounder is relevant across subgroups, its distribution may differ, leading to biased estimates of subgroup-specific treatment effects (CATEs). For example, when estimating the CATE of homeownership on unemployment among low-income individuals, if risk tolerance affects both homeownership and employment stability, and its distribution varies within this group compared to others, the estimated effect may not reflect income per se, but instead unobserved differences in risk preferences.

While researchers may not eliminate all unmeasured confounding mechanisms, a transparent approach to addressing incomplete mechanisms at each stage strengthens the credibility and persuasiveness of causal claims regarding $X \rightarrow Y$ (Manski 1990). To systematically assess these challenges, researchers can begin by constructing a causal graph (Figure 3, Step 1) to explicitly map the relationships from theory and theoretical estimands. This step serves as the foundation for sensitivity analysis, helping researchers evaluate how unmeasured confounders may distort estimand selection and shape theoretical interpretations. As we elaborate in the following five-step framework, this graphical approach provides a structured entry point for identifying potential biases and guiding subsequent sensitivity assessments.

### 2.3 Identification-assumption Sensitivity

Even if we correctly identify the main mechanisms or theoretical models, our chosen functional forms, identification strategies, or modeling assumptions may misrepresent the true data-generating process. *Identification-assumption sensitivity refers to the extent to which identification strategies rely on untestable assumptions about unmeasured confounders.* They are untestable because our social theories, and thus statistical models, try to depict the social world yet have no direct access to the mechanics of that world. If our theory wrongly misses the existence of a confounder or it knows it exists, but it remains unmeasured, then the empirical estimand does not faithfully represent the theoretical estimand.

Despite the fact that causal inference relies on untestable assumptions, the potential outcome framework (PO) and directed acyclic graph (DAG) help to clarify where these assumptions enter the analysis. These frameworks are the most widely used causal inference tools in social sciences (Gangl 2010; Morgan and Winship 2015; Pearl 2019). However, they may struggle to accommodate complexities such as multiple interventions, dynamic effects, interference, and feedback loops. For this reason, this study (or method) still relies on the concepts and assumptions of PO/DAG as the overall framework for building the analysis strategy. Through additional hypothesis testing in corresponding analysis steps (such as sensitivity analysis), we aim to mitigate the shortcomings of these idealized frameworks. When facing more complex social contexts, we also need to further expand or adjust these two frameworks to better align with actual mechanisms.

Sociologists and other social scientists care about the threat of unobserved confounding because if they can control that threat, they will be enabled to calculate causal effects from observational data and without having to run an experiment (Imbens 2023; Morgan and Winship 2015; Pearl 2009b; Rubin 1974). The unconfoundedness assumption requires that all confounders influencing both treatment and outcome are accounted for. In the framework of directed acyclic graphs (DAGs), this condition is reflected in the d-separation criterion, which ensures that all backdoor paths between treatment and outcome are blocked (Gangl 2010; Pearl 2009a; Rubin 1974).While PO and DAG frameworks establish broad assumptions about unmeasured confounders, in practice, researchers must specify the precise conditions under which causal effects can be identified from data—this process is known as identification. To address this challenge, we examine two common identification strategies used in both experimental and non-experimental settings. In fact, there is an extensive menu of designs that can be deployed to enable identification—e.g., difference-in-difference, regression discontinuity design, and instrumental variables (Angrist, Imbens, and Rubin 1996; Rubin 1974). Therefore, we show perhaps the two most common scenarios.

Using the potential outcomes framework, we define causal effect at the individual level, $i$, as the difference between two potential outcomes, $\tau_i = Y_{1i} - Y_{0i}$. Here, $Y_{1i}$ denotes the potential outcome (e.g., income at age 30) when an individual is exposed to a treatment of interest, $T$, (e.g., attending university) and $Y_{0i}$ otherwise. However, for each individual in the data collected, a sociologist will only be able to observe one of these potential outcomes, revealing only the outcome that coincides with the selected exposure status and thus rendering the second status unobserved. For example, if Jonh Dow is observed in the data and he attended university, then his observed outcome $Y$ in terms of income at age 30 coincides with his potential outcome under exposure, $Y_1$, and thus, we write $Y = Y_1$. However, that leaves his other potential outcome unobserved, for what would have happened to his income if he had not attended university. Consequently, sociologists are only able to identify one of the outcomes in observational data, leaving the causal effect $\tau_i$ unidentified, because that effect is the difference between the two potential outcomes.

Adjusting (or controlling) for confounders strategy is often used(Gangl 2010; Heckman and Pinto 2024; Pearl 2009b). Assume that parental education ($C$) and parental occupation ($X$) are the only relevant confounders. If these are observed, the conditional unconfoundedness assumption holds: $T \perp\!\!\!\perp Y_1, Y_0 \mid C, X$. This assumption allows us to treat observed outcomes as proxies for missing potential outcomes, given that units with similar values of $C$ and $X$ are comparable. Since only one of the two potential outcomes is observed for each individual (depending on exposure), we can estimate the causal effect by replacing missing counterfactuals with observed outcomes from similar individuals. Various statistical methods—such as matching, OLS, or machine learning models—can then be used to estimate the effect.

In experimental settings, Randomized Controlled Trials (RCTs) are considered the gold standard for causal inference because they use randomization to balance observed and unobserved confounders on average by design. The sociologist would ideally sample units from the entire target population and randomly assign the exposure of interest. The major advantage of randomization is that it breaks the influence of confounding, and thus, when the experiment is conducted according to all the rules of the trade (Greene 2003), When conducted in large samples and when all the units complied with the treatment assignment, the causal estimate obtained will coincide with the ATE in the observational setting, however, which still assumes that there are no unmeasured confounders that could impact compliance.

Different estimation methods impose varying assumptions about unmeasured confounders, making the choice of identification strategy context-dependent. The relationships between observed variables and unmeasured confounders, as well as alternative measurement strategies, both influences how researchers evaluate the unconfoundedness assumption. While these identification strategies aim to reduce confounding, they are built on assumptions that are often untestable (Angrist and Pischke 2009). The degree to which these assumptions hold varies across research contexts, making identification inherently uncertain. Sensitivity analysis provides a systematic approach to assessing how violations of these assumptions—particularly those related to unmeasured confounding—could impact causal conclusions. In Step 2 of our framework, we demonstrate how sensitivity analysis can be integrated into different identification strategies to quantify their robustness.

### 2.4 Specification Sensitivity

Even with well-specified models and identification strategies, uncertainty remains due to unmeasured confounders. Here, unmeasured confounders contribute to uncertainty within the model specification. *This uncertainty concerns how statistical model specifications interact with the specification of unmeasured confounding, a concept we term "specification sensitivity."*

In the move from defining empirical estimands to selecting empirical strategies, the goal is to identify a statistical model that appropriately estimates the causal effect of interest. At this stage, considerations of unmeasured confounding shift from questions of identification to how assumptions about unobserved variables are encoded within statistical models, particularly in sensitivity analysis. The way a model is specified—both in terms of its functional form and how it incorporates assumptions about unmeasured confounding—shapes the tradeoff between statistical efficiency and substantive validity. Specifically, model specification and the parameterization of unmeasured confounding jointly determine (1) the interpretation of sensitivity parameters, and (2) the structure of the model used to estimate them.

First, different metrics of sensitivity analysis parameters are based on the scales of outcome, treatment and observed variables, which impacts how we should frame sensitivity analysis. If $U$ is partially observed through a noisy proxy $W$, including $W$ in the model might reduce bias but not eliminate it. If $W$ poorly measures $U$ (e.g., dichotomizing an inherently continuous confounder), the remaining bias can be significant– an issue known as bias amplification or attenuation bias (Carroll et al. 2006). Thus, the choice to include a proxy and how it's included (as a linear control, categorical variable, etc.) will affect results. Sensitivity analysis can incorporate this by treating the proxy as part of the sensitivity parameter (Cinelli and Hazlett 2020 allow observed covariates to be factored into the parameters, for instance).

Second, if the true effect of an unmeasured confounder is nonlinear, but the model is specified as linear, the estimated treatment effect may absorb part of the confounder's influence in an unpredictable way, leading to heterogeneous estimates with unpredictable directional bias. Moreover, if unmeasured confounders moderate the treatment effect, but the model assumes a homogeneous effect, the estimated CATE may reflect a different subgroup from the intended one. These issues highlight that unmeasured confounders not only affect causal identification but also fundamentally shape how we specify statistical models, influencing the robustness, efficiency, and interpretability of causal estimates (Pearl 2009b). Thus, part of specification sensitivity is recognizing that even if $U$ doesn't bias the overall effect, it could bias who we think is experiencing the effect.

Meanwhile, It is important to distinguish between conventional "robustness checks" and the form of sensitivity analysis we emphasize (Neumayer and Plümper 2017; Saltelli et al. 2019). Robustness checks typically involve varying model specifications, functional forms, or sample subsets to see if the estimated effect remains stable– they address model uncertainty or other design choices (Saltelli et al. 2019; Young and Holsteen 2015). Sensitivity analysis, in contrast, explicitly addresses confounding uncertainty by evaluating how the effect estimate would change under different assumptions about unobserved confounders. Both are complementary: robustness checks ensure an estimate isn't an artifact of particular modeling decisions, while sensitivity analysis probes whether an estimate could be an artifact of hidden bias. The purposes of sensitivity analysis versus uncertainty analysis or robustness checks are often conflated in social sciences, which is also shown by our review of current literature within the American Sociological Review and American Journal of Sociology in Appendix B. This observation illustrates a broader methodological caution in the field, where concerns about unmeasured confounding pose significant challenges for causal inference. However, by not directly addressing the challenge of unmeasured confounding, we argue that the field may miss opportunities to generate insights that are both causally informative and sociologically relevant Additionally, our review reveals that most sociological studies employ robustness analysis without a clear definition of its purpose. Robustness checks alone are insufficient to fully address concerns about unmeasured confounding. Robustness analysis focuses on concerns of functional form, which is another methodological issue of importance discussed by other studies (see e.g., Balgi et al. 2025), Young and Holsteen (2015) and Jeong and Rothenhäusler (2022). While both robustness analysis and sensitivity analysis are required to enhance social science research (Neumayer and Plümper 2017), we focus on sensitivity analysis in this article. The step 3 and step 4 mainly explore how such parameters of unmeasured confounding and their functional forms work.

Unconfounding can influence all stages of inference– the definition of the estimand, the validity of the identification strategy, and the performance of the statistical model. By breaking the problem down this way, researchers can better pinpoint where their analysis is most vulnerable and address it. In the next section, we introduce a practical five-step workflow that embeds these considerations into the research process.

## 3. A Workflow for Assessing the Influence of the Unobserved Confounders

To bring clarity to how applied researchers may assess the influence of unobserved confounding, we developed a workflow based on our review of existing sensitivity analysis methods and our synthesis of them. Although there are many specific methods for conducting sensitivity analysis, it remains unclear when a researcher should select one method over another. The workflow aims to fill that gap, facilitating such analyses.

Our initial search identified more than 80 methodological articles, but about 30 of those articles were about robustness analysis (handling functional form issues) rather than sensitivity analysis (assessing the potential influence of unmeasured confounding). While Appendix C lists all the relevant articles and our categorization of them and shows the workflow—a synthesis of reviewing those articles. This workflow consists of five steps. Our five-step framework is designed to systematically explore the three dimensions of sensitivity in causal inference.

[Insert Figure 3 here]

*Step 1: Given your theoretical estimands, what unobserved confounding should you worry about, according to social theories?*

Each estimand is associated with specific unobserved confounders—variables that occupy a certain position in the DAG and, if unobserved, may compromise how researchers truly define target population and unit-specific quantity, as we discussed in confounding-estimand sensitivity. The purpose of discussing unmeasured confounding in this step is about exploring how to use theory to guide and frame the roles and positions of unmeasured confounding in the beginning of research choices. As noted by Hernán et al. (2002), causal knowledge rather than data is key sources of identifying bias. And, these clarifications should draw on the best available social theory (Swedberg 2017). Figure 3, in the first step, shows several estimands and their unobserved confounders, which indicates how theoretical models could be framed as DAGs. For the ATE estimand, and as previously discussed, that confounder is a common cause of the treatment and the outcome. We call such a variable treatment-outcome confounder.

While the LATE estimand implies a different theoretical model from ATE estimand, where researchers can only consider how their theory could explain effects in individuals who comply with treatment assigned to their sample group (Angrist et al. 1996). However, it is  susceptible to a distinct type of confounder—one that lies between instrument and outcome (Zhang, Stamey, and Mathur 2020). In Figure 3, the first step, we let C be the unobserved confounders that the instrument seeks to counter, but C should also not impact both instruments and outcome (the red line between Z and C). If such unmeasured confounding exists, it may cause selection into compliance to distort the target population and cause biased subgroup effect, which destroys how theory links to theoretical estimands.

Similarly, such distortion of unmeasured confounding could be found in other theoretical estimands. the Average Treatment Interaction (ATI), time-varying treatment effect, and path-specific causal effects (PCE) each have specific unobserved confounders that must be considered. When analyzing two treatments, the ATI has three such confounders: first-treatment-outcome, second-treatment-outcome, and both-treatment-outcome confounders, respectively. Figure 3 shows the ATI for a two-treatment situation but the ATI may have several treatments of interest making it a multi-treatment estimand (Keele and Stevenson 2020; Zheng, D'Amour, and Franks 2025). When there are multiple treatments, scholars need to consider potential confounding for each treatment-outcome and how that confounder groups with other treatments and the outcome. The time-varying treatment effect and the PCE have a similar confounding structure to the ATI but now that causal effect is unfolding over time or a specific path[1].

Thus, the ATE, ATI, and PCE share similar challenges related to confounding but differ in specific ways. Using a cross-sectional design, the ATE faces one set of confounding, while the ATI faces two sets referring to the two treatments. If the two treatments have the same confounding set, then they are identical to the ATE set up. The PCE on the other hand, faces a set of confounding that are common causes of different combinations: treatment-outcome, treatment-mediator, and mediator outcome, making the confounding problem more complicated. Therefore, we could find that each theoretical estimand implies different theoretical paths with various unmeasured confounders.

*Step 2: What observed confounding, proxies, or drivers can you include in the adjustment set, to reduce causal bias?*

Step 1 explore how theory could be framed as DAGs to present theoretical estimands with the discussion on unmeasured confounding (Daoud and Dubhashi 2023). In conjunction with the first step, scholars need to clarify what confounding is observed and what will remain unobserved, despite all available research efforts, while considering practical constraints related to data availability. Step 2 indicates how we should consider the real adjustments, identifications and assumptions from theoretical estimands to empirical estimands, which mainly discuss identification-assumption sensitivity in empirical settings. Adjusting for observed confounding is not only necessary for identification but such adjustment may alleviate the influence of the unobserved confounders, assuming that these confounders correlate with each other (VanderWeele 2019).

Besides adjusting for observed confounding, proxies or drivers included in the adjustment set may reduce causal bias, because such inclusion may alleviate the influence of unobserved confounding (Pearl 2013). A proxy is a variable that is affected by the confounder; a driver is a variable that affects the confounder (Pearl 2013; Sakamoto, Jerzak, and Daoud 2024). The intuition for why a proxy or a driver alleviates unobserved confounding is that they contain informational traces of the confounding. While a proxy is injected with such traces by the confounder, the driver injects information into the confounder—represented by the arrow pointing from the driver to the unobserved confounder. While rare, an ideal driver or proxy would capture most relevant information about the confounder. Even an imperfect proxy or driver can still be useful in reducing the influence of unobserved confounding. Furthermore, the conditional permutation test could be used for independence while controlling for confounders (Berrett et al. 2020).

For a proxy to be effective and minimize bias attenuation, it should closely reflect the hidden confounding variable it represents. That means that it should accurately represent the statistical signal in that original variable. For example, if the proxy is a dichotomized representation of a continuous confounder, the proxy may fail to accurately represent that continuity, thereby, resulting in measurement error. That error translates to attenuation bias (Gabriel, Peña, and Sjölander 2022).

Some sensitivity analysis methods enable a readily available procedure to include observed confounding, proxies, or drivers. Cinelli and Hazlett (2020) provide such a possibility. Typically, parameters for sensitivity analysis are set together with observed confounding variables. In simulation-based methods, these observable confounders are incorporated into the function that simulates the effect of the unmeasured confounder. For example, Bijlsma and Wilson's (2020) approach simulates the unmeasured confounder using only treatment and outcome variables, whereas Carnegie, Harada, and Hill (2016) incorporate observable confounders into their simulation. Once researchers make the best effort to framework their estimands, step 3 and step 4 are used to explore how statistical model specifications and specifications of unmeasured confounding jointly impact the estimation strategies.

*Step 3: What is the scale of the outcome, treatment, and unmeasured confounder, and what sensitivity parameter is most interpretable for that scale?*

The scale of unobserved confounding, treatment, and the outcome influences what sensitivity analysis method one can use (Rene, Linero, and Slate 2023). For example, if the unobserved variable of interest is biological sex, then the scale is binary; if it is gender, then the variable is nominal, containing several categories; if the variable is income, it is continuous. Thus, depending on the scale of this variable, scholars may make different assumptions on that confounding. Although most methods accommodate various scales of unobserved confounders, some impose constraints on the scale of the treatment and, occasionally, the outcome (Peña's 2022). Binary scales are commonly used due to their relative simplicity in certain methods (Greenland 1996; Lindmark, de Luna, and Eriksson 2018; Peña 2022; Rosenbaum and Rubin 1983; Sjölander and Hössjer 2021).

Depending on the scale of interest, scholars may use different sensitivity parameters to characterize their analysis. A sensitivity parameter is a statistical metric that encodes the assumed strength of confounding—and potentially its strength with observed variables. The sensitivity analysis literature provides several types of sensitivity parameters, as shown in Step 3, Figure 3. Often, most tools provide two parameters: the first enables the analyst to control the assumed statistical magnitude between the confounder and treatment, and the second enables control over the confounder-outcome magnitude. Because these relationships are not directly observed, sensitivity analysis involves both evaluating hypothetical magnitudes—simulating what would have happened if the scholar had adjusted for the confounder—and assessing the actual statistical signal (covariance) between the treatment and outcome. As this signal strengthens, the treatment-outcome relationship becomes more robust to hypothetical unobserved confounding.

In principle, these magnitudes can be encoded in any type of statistical measure, but the most common are relative risks (also known as risk ratio) (VanderWeele and Ding 2017), linear parameters (Carnegie et al. 2016; Dorie et al. 2016), coefficient of determination (commonly known as $R^2$)(Cinelli and Hazlett 2020), and correlations (Balgi et al. 2024). The choice of metric should be informed by the scale of the variables included in the sensitivity analysis, as well as substantive interpretability.

Step 3, Figure 3, shows a case when the outcome and treatment are binary. In such cases, probability distributions may provide a more intuitive way to assess the influence of confounding. A probability distribution encodes how two or more variables depend on each other in terms of probabilities. The E-values method uses probability distributions to construct relative risks (also known as risk ratios) to reason about the effect of unobserved confounding. This method can also use risk difference Sjölander and Hössjer (2021), hazard ratio, or odds ratios (VanderWeele and Ding 2017), to achieve the same goal. Relative risk (RR) is the ratio between two probability distributions. The main interest is often the ratio of probability between the outcome of interest occurring in the treated group divided by the outcome occurring in the control group, $RR_{TY} = \frac{P(Y=1|T=1)}{P(Y=1|T=0)}$.

In the setting of relative risk, the user must reason about two sensitivity parameters. The first parameter is the risk ratio of $RR_{TC} = \frac{P(C=c|T=1)}{P(C=c|T=0)}$, which tells us how imbalanced the treatment groups are in the unmeasured confounder. For example, if $80\%$ of youth attending university also had parents with higher education, as compared to $40\%$ of those youth not attending university, we would have $RR_{TC} = 2$. Here, one postulates plausible values on the unobserved confounder, such that it maximizes $RR_{TC}$, and hence, why we use the operator $max_c$ in $RR_{TC} = max_c \left[\frac{P(C=c|T=1)}{P(C=c|T=0)}\right]$. Often, the max is selected to define the worst-case scenario bound if the analyst lacks prior knowledge about the strength of confounding.

Then, the analyst postulates plausible values for, $RR_{CY} = \frac{P(Y=1|C=c)}{P(Y=1|C=c^{*})}$, which defines a maximum for how important this confounder is for the outcome across levels of the confounder, $c$ and $c^*$ . For example, one might argue that a youth's future income highly depends on their parent's education level, and thus, youths with a high income in adulthood and parents with high education attainment have a 90% probability of attaining high income, compared to 30% probability if their parents do not attain higher education. Then $RR_{CY} = \frac{0.9}{0.3} = 3$. Then, the bias factor is $B = \frac{RR_{CY}RR_{TC}}{(RR_{TC}+RR_{CY}-1)}$. With our postulated magnitudes, one obtains a bias factor of $B = \frac{3 \cdot 2}{(3+2-1)} = \frac{3}{2}$ . The bias factor tells us how much the unmeasured confounding changes the obtained estimate, would have been observed with the postulated strength. We obtain that change by dividing the estimated risk ratio by the bias factor. For example, if attending university would produce a risk ratio estimate of 2, that is youth that attend university have two times the probability of getting high income in adulthood than those youth that did not attend university. Then, we divide $\frac{2}{\frac{3}{2}} = \frac{4}{3}$ and because that number is higher than one, it is not sufficient to explain away the estimated effect of 2. These E-value calculations show the magnitude in $R_{CY}$ and $R_{TC}$ such that the estimated risk ratio can be explained away, postulating a variety of magnitudes.

Probability distribution is one way of characterizing the relationship between the unmeasured confounder, treatment, and outcome. Using a parametric approach is another way. Bijlsma and Wilson (2020) develop such an approach. The more flexible framework is to combine probability distributions and linear parameters in simulations (Carnegie et al. 2016; Dorie et al. 2016). Like the relative risk approach, the researcher is expected to theorize about and stipulate plausible strengths for the unmeasured confounding. But instead of expressing that strength in relative risk, the researcher expresses it in two linear-model parameters. While one parameter, $\beta_{TC}$, expresses how strong the relationship is between the treatment and confounding, the other parameter $\beta_{CY}$ expresses the strength between the outcome and confounding. Although this sensitivity method assumes a linear functional form, commonly used modeling approach in social science research.

Continuing along the lines of familiarity, Imbens (2003) used partial R square to provide more interpretable results based on a comparison between partial R square of unobservable covariates and observable covariates. Cinelli and Hazlett (2020) developed that approach further encoding the sensitivity parameter in terms of partial explained variance (coefficient of determination), that is, in the partial $R^2$. This approach focuses on quantifying how much an unobserved confounder could account for the observed results. One advantage of this approach is that it focuses on describing the smallest statistical association needed between the unobserved confounder and both the treatment and outcome to alter the estimated effect. Recently, Chernozhukov et al. (2022) generalized this approach to any nonparametric approach, including causal machine learning. Cinelli and Hazlett (2020) encapsulate their approach and Cinelli et al. (2020), facilitate usage.

A related approach to the partial $R^2$ and the linear-parameter-$\beta$ approach, is to conceptualize the unobserved confounding as correlations (Balgi et al. 2024; Imai, Keele, and Yamamoto 2010; Lindmark et al. 2018). For example, (Balgi et al. 2024) use a correlation approach, focusing on the error in the treatment $\epsilon_T$ and outcome $\epsilon_Y$. An equivalent way of conceptualizing unobserved confounding is to say that those two errors are correlated. That is, there are one or several unobserved confounders that make those errors correlated. In the Balgi et al. (2024) approach, the researcher is asked to stipulate a set of plausible correlations, called $\rho$. One advantage of the correlation approach is that it provides an intuitive way of characterizing the statistical relationship between unobserved confounding, treatment, and outcome.

As previously mentioned, the appropriateness of the metrics chosen by the choice of metrics depends largely on the scale of the unobserved confounder, treatment, and confounder. If they are all binary, then reasoning about their relationship is more appropriate using relative risk. If they are all continuous, then correlation in error provides the most intuitive interpretation. If they include a mix of binary, nominal, and continuous variables, a parametric or variance-explained approach is often a suitable choice.

*Step 4: Selecting the Functional Relationships*

The last step is to choose a functional form for characterizing the statistical relationship among the unobserved confounding, treatment, and outcome. There are three broad classes to select from: parametric, semi-parametric, and non-parametric functional forms. The selection of an appropriate functional form is guided by an understanding of the underlying data structure and research objectives (Dorie et al. 2016). Parametric models, characterized by their reliance on a predefined functional form and a finite set of parameters, offer simplicity and ease of interpretation but may be sensitive to model misspecification if the chosen form poorly represents the data. Many of the existing sensitivity methods rely on parametric modeling (Bijlsma and Wilson 2020; Cinelli and Hazlett 2020; Imbens 2003).

Nonparametric models, which do not rely on predefined functional forms, offer greater flexibility in capturing complex relationships within large datasets but can pose challenges related to interpretability and computational efficiency. Balgi et al. (2024) developed a nonparametric model that uses a deep learning approach to estimate the full joint and factorized distribution of a stipulated DAG—the postulated causal order among the treatment, outcome, mediators, and other observed confounding. In other words, given a social theory encoded as a DAG, the deep learning model estimates the distribution from available data. Then, based on the correlation-based sensitivity parameter, one can simulate a range of values to assess when and how the obtained results might change. The E-value is also a form of nonparametric approach (VanderWeele and Ding 2017), as this approach is model and distribution agnostic.

Semiparametric models incorporate parametric assumptions for certain components while employing nonparametric techniques for others, providing a flexible approach that can address some of the limitations associated with purely parametric or nonparametric methods. The balance between flexibility, efficiency, and interpretability highlights the importance of model selection in sensitivity analysis, as the choice among parametric, nonparametric, or semiparametric models shapes how findings are interpreted (Nabi et al. 2024; Zhang and Tchetgen Tchetgen 2022).

*Step 5: Interpret sensitivity analysis outputs*

Although each sensitivity method tends to focus on one estimand, sensitivity parameter, and modeling assumption, they all share a common goal: evaluating how the quantity of interest changes, whether it moves towards zero, shifts signs, or falls within certain bounds. Sensitivity analysis has some similarities to manual stepwise regression, a regression approach where variables are sequentially added to examine how the estimate of interest changes. However, in sensitivity analysis, rather than testing only observed covariates, one assesses the potential influence of one or a set of unobserved variables. In the next section, we provide an empirical illustration demonstrating how interpretability is applied in practice.

In addition to its traditional role in identifying methodological vulnerabilities, sensitivity analysis also actively contributes to theory building and hypothesis generation in sociology. By systematically evaluating how hypothetical unobserved factors could alter causal estimates, sensitivity analysis offers detailed insights into alternative explanations and underlying mechanisms through three distinct roles: confounding-estimand sensitivity helps researchers explore the robustness of causal effects to unmeasured confounders; identification-assumption sensitivity prompts critical reconsideration of the core assumptions underpinning causal identification; and specification sensitivity encourages rigorous examination and refinement of model specifications. Thus, sensitivity analysis transcends its defensive function of merely highlighting methodological limitations by explicitly guiding sociologists toward targeted theoretical and empirical investigations.

## 4. Empirical Illustrations: Applying the Sensitivity Analysis Blueprint on the Blau and Duncan 1967 Study

We apply our blueprint for sensitivity analysis on the renowned Blau and Duncan 1967 study, as a worked example. This study explored the links between educational achievement and job status, drawing on the 1962 "Occupational Changes in a Generation" (OCG) survey. It analyzed data concerning roughly 20,000 American men aged between 20 and 64. The dataset from Blau and Duncan (1967) comprises five variables: Son's Educational Attainment ($T$), Father's Educational Attainment ($V$), Son's Socioeconomic Status at First Job ($W$), Father's Socioeconomic Status ($X$), and Son's Socioeconomic Status in 1962 ($Y$), and Unmeasured confounders ($U$). The causal relationships among these variables are shown in Figure 4. Our objective is not to duplicate the entire study by Blau and Duncan (1967) but rather to leverage their setup and dataset—we use Balgi et al. (2024) curated data. That will enable us to demonstrate the application of various sensitivity analysis techniques in two ways: one where the estimated is the average treatment effect (ATE), and the other where that estimand comprises path-specific causal effect (PCE) (Appendix D).

[Insert Figure 4 here]

### 4.1 Average Treatment Effect

The first and second steps in our sensitivity analysis workflow are to define the estimand, observed and unobserved confounding. One of the goals of the Blau and Duncan (1967) study is to analyze the average treatment effect of a son's educational achievement on his socioeconomic status in 1962, as indicated by the blue line in Figure 4. However, an observational analysis that aims to empirically estimate that estimand faces several threats, of which only some we can account for. As shown in the theoretical DAG model, the father's educational attainment and the father's socioeconomic status are observable confounders. However, the data lacks measurements of the mother's socioeconomic status, their extended family, and friends' status—factors that may likely affect the son's socioeconomic status later in life. Thus, our analysis would benefit from a sensitivity analysis to examine the potential impact of these unmeasured confounders, as indicated by the red lines in Figure 4.

Additionally, these data lack potential proxies or drivers that may function as a stand-in for the unmeasured confounders. For example, neighborhood-level socioeconomic status of where households are located might provide indirect information on the extended family and their friend's socioeconomic status. Such proxies or drivers would have reduced the bias, but they are lacking in the dataset.

In the third step we reflect on how to parametrize our sensitivity analysis. By parametrize, we mean how to operationalize the influence of the unobserved confounding. That operationalization depends to some extent on the scales of the treatment, outcome, and the unobserved confounding. In the Blau and Duncan study, the scales of outcome and treatment are both continuous, and thus, there are multiple sensitivity analysis methods that could be used. We will show two popular methods that use different parametrization: Cinelli and Hazlett (2020) and VanderWeele and Ding (2017). By contrasting these two methods, we will show how the choice of sensitivity method affects our interpretation of the unmeasured confounders on our ATE. While the main parameter of interest is explained variation in Cinelli and Hazlett (2020), the main parameter in VanderWeele and Ding (2017) is the ratio of confounding discussed earlier.

The fourth step is what functional form assumptions one Is inclined to make. Assuming a parametric form—often linearity—one gains interpretability but at the cost of omitting to model complex (interactive) relationships that may exist among the treatment, observed confounding, and outcome. For the two method choices of Cinelli and Hazlett (2020) and VanderWeele and Ding (2017), we can assume distribution-free unmeasured confounders.

In the fifth step, we interpret how strong unmeasured confounding must be to overthrow our result. That overthrowing translates to how fast our estimated causal effect may approach zero and shift signs. Starting with Cinelli and Hazlett (2020), we investigate potential breaches of the assumption that there are no unmeasured confounders in the Blau and Duncan (1967) study. This analysis aims to evaluate the ATE of a son's educational attainment ($T$) on the son's socioeconomic status in 1962 ($Y$) while accounting for the educational and socioeconomic statuses of the father. The estimated ATE is 0.32, which means on average, one category of the son's educational attainment results in son's socioeconomic status in 1962 that is 0.32 units different from the reference category. Figure 5 shows a contour map of the increasing strengths of unmeasured confounders which culminate by overthrowing our estimated result (check Appendix E to find codes and a top model for unmeasured confounders).

A counter map is the usual way of depicting sensitivity analyses. The horizontal axis represents the hypothetical proportion of variation in the treatment explained by unobserved confounders, $R^2_{T\sim U|X}$. The vertical axis illustrates the hypothetical partial $R^2$ between unobserved confounders and the outcome, $R^2_{Y\sim U|X,T}$ (Cinelli and Hazlett 2020). The contour lines on the graph indicate the estimated effect on Son's educational attainment that would result from incorporating unmeasured confounders, which could possibly be family and friend networks, or even genetic confounding (Harden and Koellinger 2020; Liu 2018). We read the contour map from the black triangle, from the bottom left up to the upper right corner, the red diamond. The black triangle assumes no explained variance in either treatment or unobserved confounding, resulting in an estimate of 0.32 We can then jump to the first red diamond, which is our estimated effect where there is no unobserved confounding by assumption. We jump to the second red diamond, and there we assume that the unobserved confounding is as strong as the observed confounding. That lightest scenario would reduce the ATE from 0.32 to 0.292. If we increase the strength of confounding to two times the strength of observed confounding, we will obtain an even further reduced estimate, down to 0.259. We have to increase the strength of unobserved confounding to about 7 times stronger than the observed Father's socioeconomic status to hist zero and overthrow the analysis (Cinelli and Hazlett 2020). Beyond this strength, the analysis shifts sign and the original interpretation of the ATE completely shifts.

Given that the observed confounding—father's education—is likely a strong common cause of the son's socioeconomic conditions, our interpretation is that unobserved confounding has to be quite strong to pose a threat to the Blau and Duncan original conclusions.

[Insert Figure 5 here]

Let us re-run the sensitivity analysis but now using the E-value method, based on VanderWeele and Ding (2017). This analysis uses linear regression for analyses involving continuous exposures and outcomes in this case. To align with this approach, the function implements effect-size conversion techniques as detailed by Chinn (2000) and VanderWeele (2017), specifically designed for continuous variables. These techniques aim to derive an approximate odds ratio by converting the continuous outcome into dichotomous categories, based on the mean difference between exposure groups, as discussed in VanderWeele and Ding (2017). Based on that, we use the average treatment effect (ATE=0.32) with effect-size conversion techniques developed by Chinn (2000) and VanderWeele (2017), to get the RR that is about 1.34, indicating the risk of the outcome occurring in the exposed group is 1.34 times than the risk of the outcome occurring in the control group.

[Insert Figure 6 here]

Similar to the previous analysis, we use contour maps, shown in Figure 6. Here, the y-axis is the ratio $RR_{UY}$, which reflects the significance of the unobserved confounder in influencing the son's socioeconomic status in 1962. That ratio captures the outcome differences due to unmeasured confounders across all treatment groups within each propensity-score stratum (VanderWeele and Ding 2017). While the x-axis, the ratio $RR_{TU}$, captures the extent of imbalance between the treated group and the untreated group regarding the unobserved confounder. That ratio quantifies the difference in prevalence of unmeasured confounders between the treated group and the untreated group (VanderWeele and Ding 2017). For instance, one possible unmeasured confounder could be the mother's education. Therefore, $RR_{UY}$ reflects the significance of the unmeasured mother's education in influencing the son's socioeconomic status in 1962. And $RR_{TU}$ captures the extent of imbalance between the treated group and the untreated group regarding the mother's education.

Here, as proved by Ding and VanderWeele (2016), the true relative risk $RR_{TY}^{true}$ must be at least as large as the $RR_{TY}^{obs}/\frac{RR_{TU}RR_{UY}}{(RR_{TU}+RR_{UY}-1)}$, and $\frac{RR_{TU}RR_{UY}}{(RR_{TU}+RR_{UY}-1)}$ is the joint bounding factor based these two sensitivity parameters, which is relevant to the bound of the true effect in the presence of unmeasured confounders (Ding and VanderWeele 2016). $RR_{TY}^{true} = 1$ indicates no difference between the treated group and the untreated group, in another word, the true effect is null. Meanwhile, $\frac{RR_{TU}RR_{UY}}{(RR_{TU}+RR_{UY}-1)}$ should be equal to $RR_{TY}^{obs}$ to make $RR_{TY}^{obs}/\frac{RR_{TU}RR_{UY}}{(RR_{TU}+RR_{UY}-1)}$ to be 1. Therefore, the idea behind the E-value is to find a size for both $RR_{TU} and\ RR_{UY}$ to make $\frac{RR_{TU}RR_{UY}}{(RR_{TU}+RR_{UY}-1)}$=$RR_{TY}^{obs}$. In our case, the $RR_{TY}^{obs}$ is about 1.34, and when $RR_{TU} = RR_{UY} = 2.01$, $\frac{2.01*2.01}{(2.01+2.01-1)} \approx 1.34$. Therefore, if the outcome difference due to the mother's education across all treatment groups within each propensity-score stratum is 2.01 and the difference in prevalence of the mother's education between the "high" educational attainment group and the "low" educational attainment group is also 2.01, the observed association between the son's educational attainment and the son's socioeconomic status in 1962 (RR = 1.34) might be entirely explained away by the unmeasured mother's education.

As shown in Figure 6, an E-value of 2.01 implies that the unmeasured confounders must at least double the probability of an individual education attainment in the treatment and control. That means that an unmeasured confounder needs to double the probability across levels of educational attainment. As previously mentioned, in our case, we had to dichotomize the exposure variable and thus our comparison of treated and control along the chosen threshold (VanderWeele 2017). Moreover, the confounders would be required to similarly double the likelihood of an individual being classified as "high" rather than "low" regarding the outcome, which indicates the two-fold impacts of the unobserved confounder in influencing the son's socioeconomic status in 1962. The decision to categorize outcomes as "high" or "low" is made theoretically and relies on specific distributional assumptions outlined in both Chinn (2000) and VanderWeele (2017).

### 4.2 Summarizing the steps

In using the framework, we make the following reflections on the state of current methodologies. First, even using the same data, the definition of ATE slightly varies with the scale of the treatment. For example, when the treatment is continuous, the causal dose-response function is often considered as the estimand, although it is equivalent to the average treatment effect given the treatment level.

Second, in step 2, it is crucial to pinpoint where observed confounders lie. In the Blau and Duncan study, the father's educational level and socioeconomic status act as pre-treatment confounders. Yet, accounting for post-treatment confounders (like Son's Socioeconomic Status at the first job) could skew the results. Additionally, in certain scenarios, what's considered post-treatment confounders could also serve as a mediator, depending on the specific theoretical estimand being pursued.

Third, various sensitivity analysis approaches apply different scales to their sensitivity parameters. In our scenario, treating the son's educational attainment as a categorical variable may not align perfectly with Cinelli and Hazlett's (2020) framework, which does not detail evaluating sensitivity parameters across categories. The E-value approach also encounters a scaling challenge, as it must translate continuous outcomes into a risk ratio scale, potentially losing information and affecting its distribution.

Fourth, while the assumptions underpinning the two sensitivity methods, we used to be fairly adaptable, they might lead the reader to make some implicit hypotheses. For example, the method developed by Cinelli and Hazlett (2020), only requires maximum explanatory power of unmeasured confounders.

## 5. Discussion

The importance of handling unobserved confounding cuts through virtually all social science research (Oster 2019; Rosenbaum 2002a; Van Der Weele and Ding 2017).. Our framework suggests a more proactive approach: rather than avoiding causal claims due to unobservables, we encourage integrating sensitivity analysis to openly grapple with the uncertainty they introduce. By doing so, we can better gauge how much hidden bias would be needed to undermine our conclusions and communicate that to our audience. The framework, centered around applying sensitivity analysis, offers an approach that not only acknowledges but actively engages with the complexities of causal inference in observational studies. It also encourages improved data on potential unobserved variables as well as actively refining social theory on the role of such variables. Sensitivity analysis, as we elaborate, is not merely a technical exercise but a critical methodological tool towards more robust and transparent sociological research. It equips sociologists to answer the question, 'What would it take for this result to no longer hold?' Moreover, our empirical illustration, leveraging the Blau and Duncan study, serves as a concrete example of how sensitivity analysis can be pragmatically applied in sociological research—in addition to the exemplary sociological studies discussed in Section 5. Similarly, Wodtke et al. (2016) used sensitivity analysis to show that neighborhood effects on incarceration persisted unless selection bias was implausibly large. This illustration not only demonstrates the feasibility of incorporating sensitivity analysis into sociological studies but also highlights its potential to clarify and strengthen causal claims.

The adoption of sensitivity analysis represents more than a methodological refinement—it marks a shift in how sociologists engage with causal claims. Moving from certainty-seeking to transparency-driven reasoning (Hedstrom 2005), sensitivity analysis encourages explicit reflection on the assumptions that underlie our inferences. Rather than treating unmeasured confounding as a nuisance to be ignored or bracketed, it offers a systematic approach to assessing how such uncertainties might influence causal conclusions. This reorients causal inference from a binary of "identified or not" toward a continuum of credibility, grounded in both methodological rigor and theoretical clarity. Such reasoning further demands attention to context-dependence, recognizing that causal effects may vary across time, space, and subpopulations (Hernán et al. 2002).

Beyond strengthening individual studies, the broader adoption of sensitivity analysis promotes a disciplinary culture of explicit assumption-checking. If multiple studies in a field demonstrate that results hold despite plausible confounding, collective confidence in the causal claim increases—even in the absence of definitive proof. Conversely, when sensitivity analyses suggest fragility, researchers can transparently signal the need for better data or more rigorous design. In this sense, sensitivity analysis is not merely a diagnostic tool, but a mechanism for accumulating sociological knowledge under conditions of uncertainty.

To enhance the uptake and continued development of sensitivity analysis, methodologists have an opportunity to consider unmeasured confounders from a theoretical perspective to an estimation-level tool. Based on our review, we find that there is an increasing methodological specialization, perhaps hindering applied researchers from grasping all the technical nuances. For example, several of the methods discussed lack software, as shown in the Appendix C. Additionally, while current sensitivity methods tend to specialize in targeting specific estimands, there have been few attempts to unify existing methods under a wider umbrella of a method. For example, while Balgi et al. (2024) offer a way to target virtually any estimand of interest, it is limited to using a correlation metric as a sensitivity parameter, missing other parameters such as explained variance (Cinelli and Hazlett 2020). Another avenue for methodological development is to enable sensitivity analysis over two or more unobserved confounding. Current methods tend to assume that applied researchers focus on one confounding variable or can summarize all confounding as a set of confounding variables (i.e., call it context). They can reason about the average influence that exhibits over the treatment and outcome. However, often applied researchers may want to disentangle that into two or more subsets, where they can reason about the influence of context more carefully. Such a sensitivity method is currently lacking but entirely possible to develop statistically.

Despite emphasizing the importance of sensitivity analysis in causal inference for the social sciences, it is essential to acknowledge its limitations. First, sensitivity analysis relies on researchers' assumptions about unmeasured confounders, such as their strength, distribution, and relationship with both the treatment and outcome variables (Cinelli and Hazlett 2020). These assumptions are often based on theoretical reasoning or previous literature, yet they remain subjective and may not fully capture real-world confounding mechanisms (Rosenbaum 2002b).

Second, the robustness of sensitivity analysis is constrained by model specifications. Many sensitivity analysis methods assume that the effect of unmeasured confounders is linear, whereas real-world causal relationships are often more complex, involving nonlinear effects, interactions, and high-dimensional confounders (Greenland 2003; VanderWeele and Ding 2017). When the model specification deviates from the true data-generating process, the results of sensitivity analysis may be misleading.

Furthermore, sensitivity analysis does not directly address causal identification but instead assesses the potential impact of unmeasured confounders on estimated results (Imbens 2003). Thus, even if a causal estimate appears “robust” under certain assumptions about unmeasured confounders, it does not guarantee that the causal interpretation is correct. For instance, some unmeasured confounders may have asymmetric effects, meaning their influence is more pronounced in certain subgroups. Sensitivity analysis by itself does not address other threats to causal inference besides confounding. Selection bias (collider bias) due to conditioning on certain variables, and measurement error are examples of issues requiring their own assessments.

Our article calls for methodological reorientation within sociology, advocating for a more explicit and structured engagement with the challenges of causal inference. Defining your estimand is one critical step (Auspurg and Brüderl 2021); evaluating how unmeasured confounders impact the critical choices in quantitative sociology is another. By embracing this approach, sociology can move towards a more causally informed and methodologically transparent future, enhancing its capacity to contribute to a deeper understanding of the social world.

**Notes**

1. There are multiple estimands on mediating roles, but we do not discuss the differences between them, please check (VanderWeele 2016).

## Figures

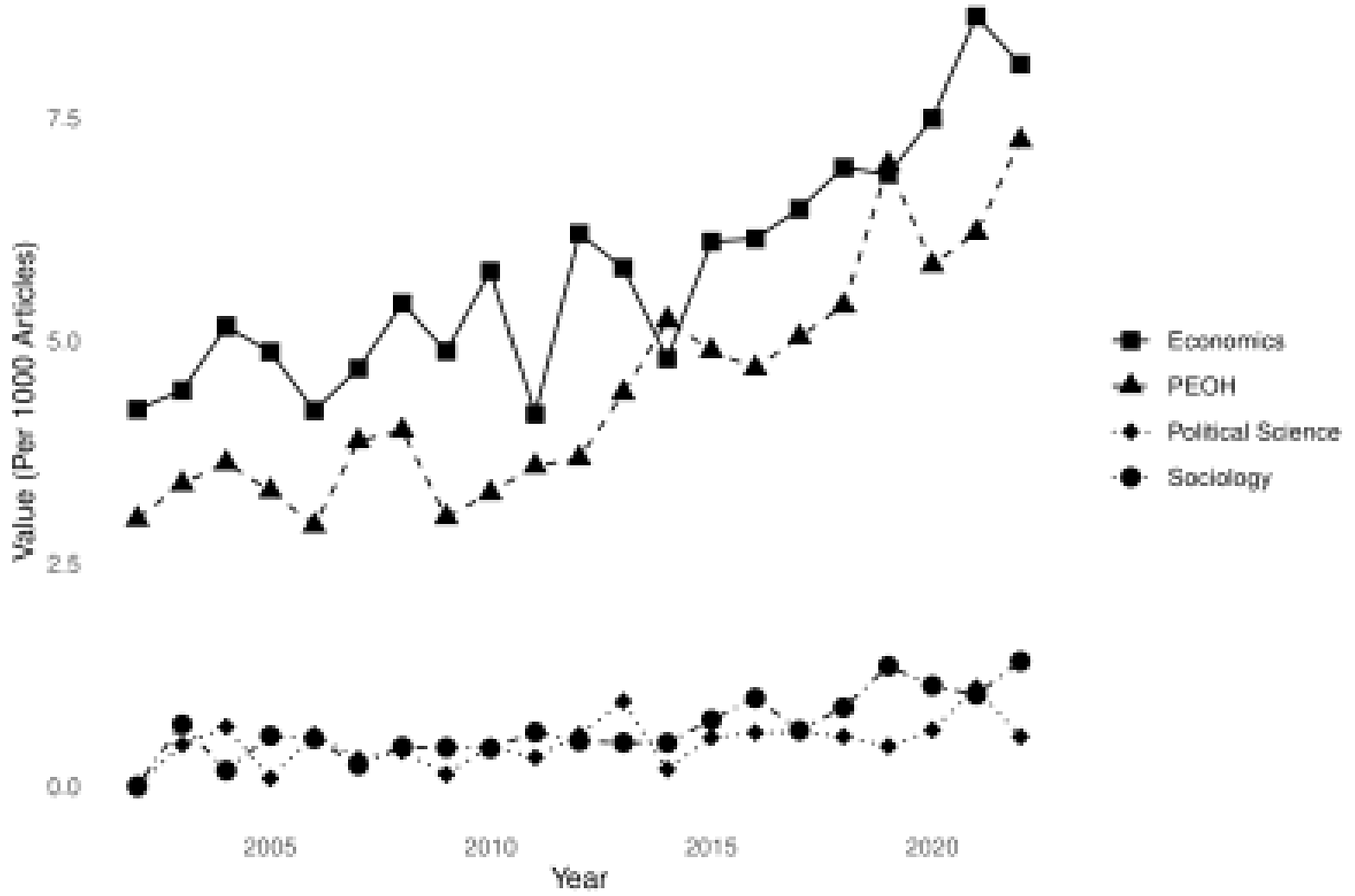

**Figure 1.The Prevalence of Usage of "Sensitivity Analysis" and its Relevant Terms by Disciplines, 2002 to 2022**

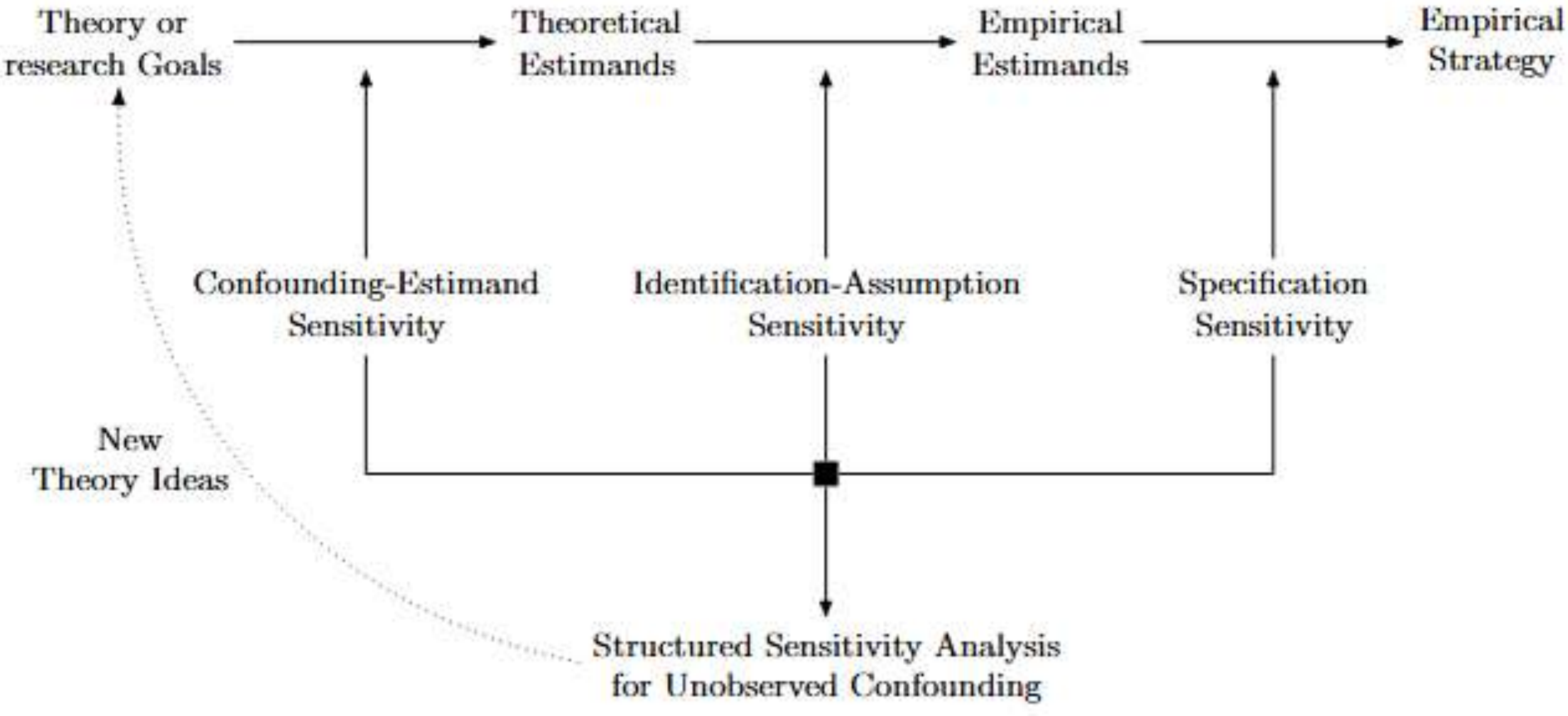

**Figure 2. The Sensitivity of Statistical Analysis in Sociology**

**Step 1. Estimands and Unobserved Confounders**

What is your estimand and where is the unobserved confounder?

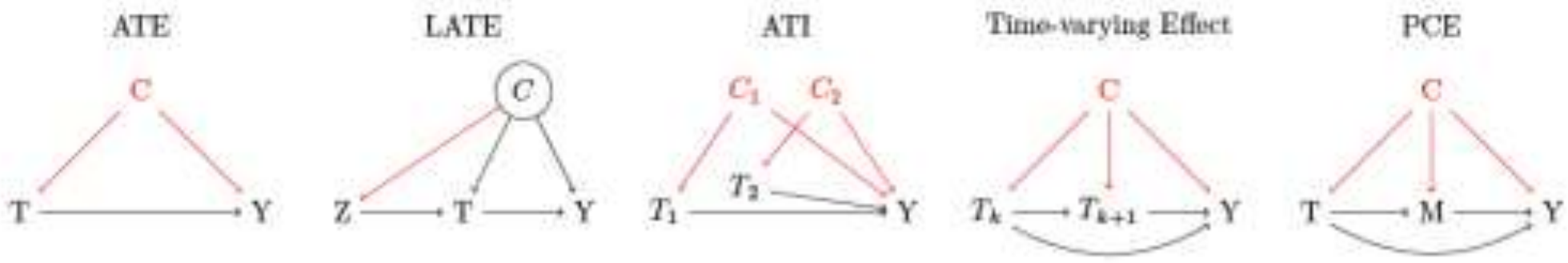

**Step 2. Adjust for Observed Confounders, Proxies and Drivers**

Are you adjusting for observed confounders $X$, proxies $P$ or drivers $D$ of unobserved confounders ?

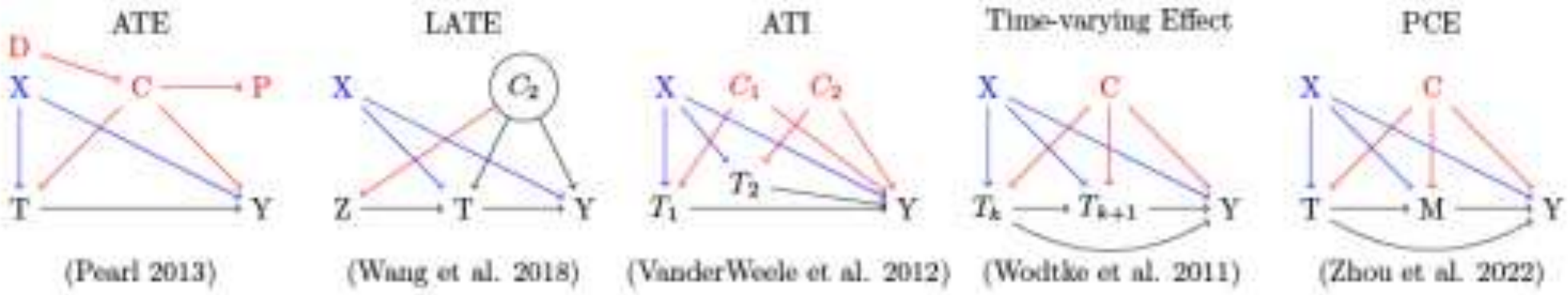

**Step 3. The Parameters of the Unobserved Confounder**

What are the scales of the treatment, the outcome and the unobserved confounder, and what is an appropriate metric in quantifying the influence of the unmeasured confounder?

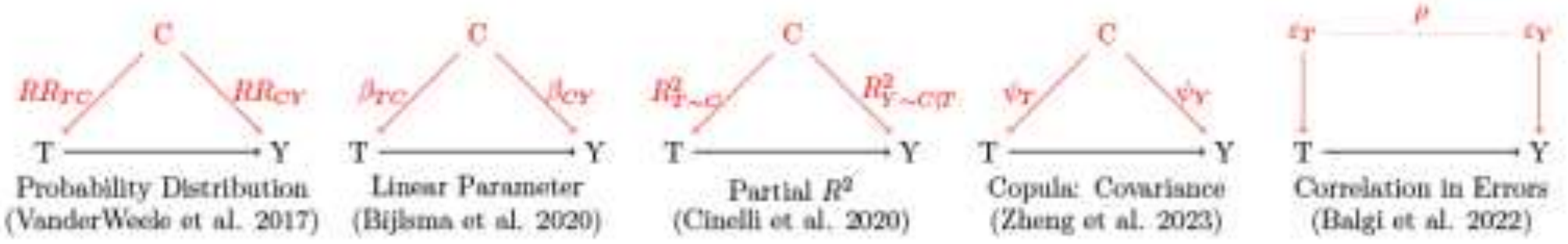

**Step 4. Functional Assumptions**

What functional form captures the relationship among the treatment, outcome, and confounders?

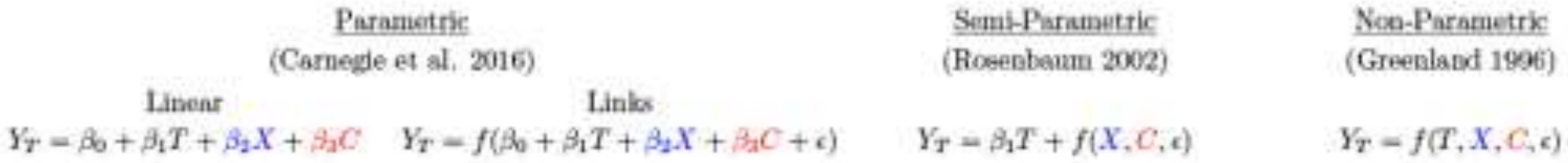

**Step 5. Interpretation**

In this last step, one analyzes how fast the causal quantity moves toward zero , shifts sign, or falls within bounds.

**Figure 3. The Blueprint of Workflow for Sensitivity Analysis**

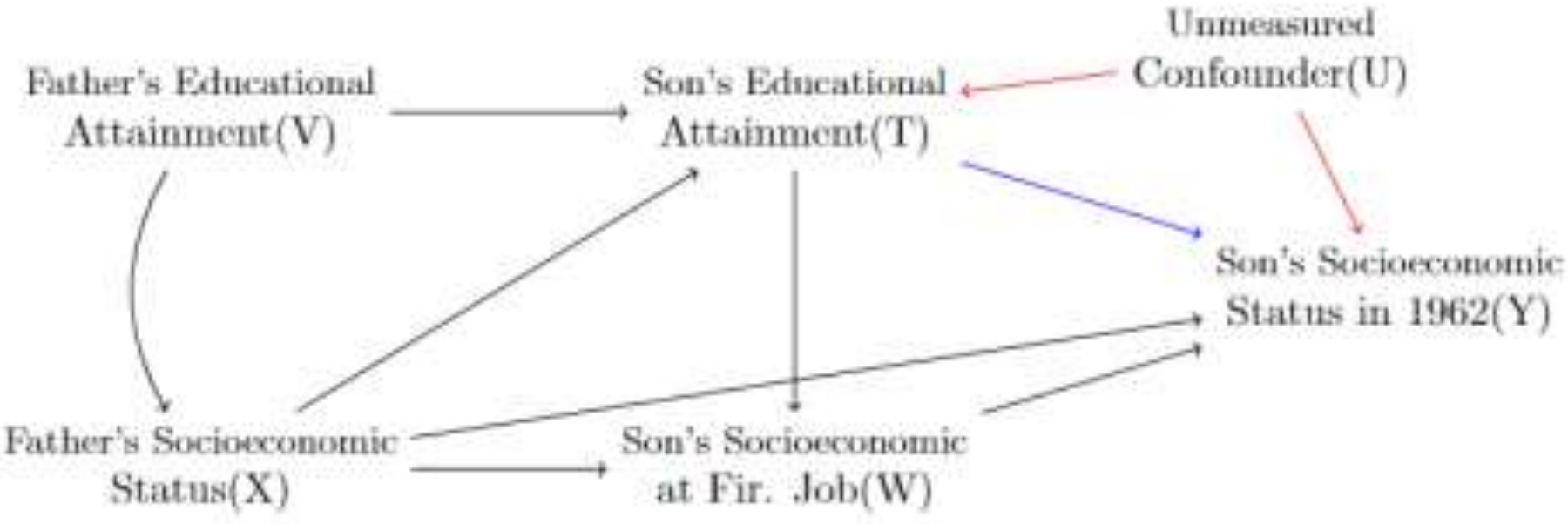

**Figure 4. A DAG Corresponding to Blau and Duncan's (1967) Linear Path Model**

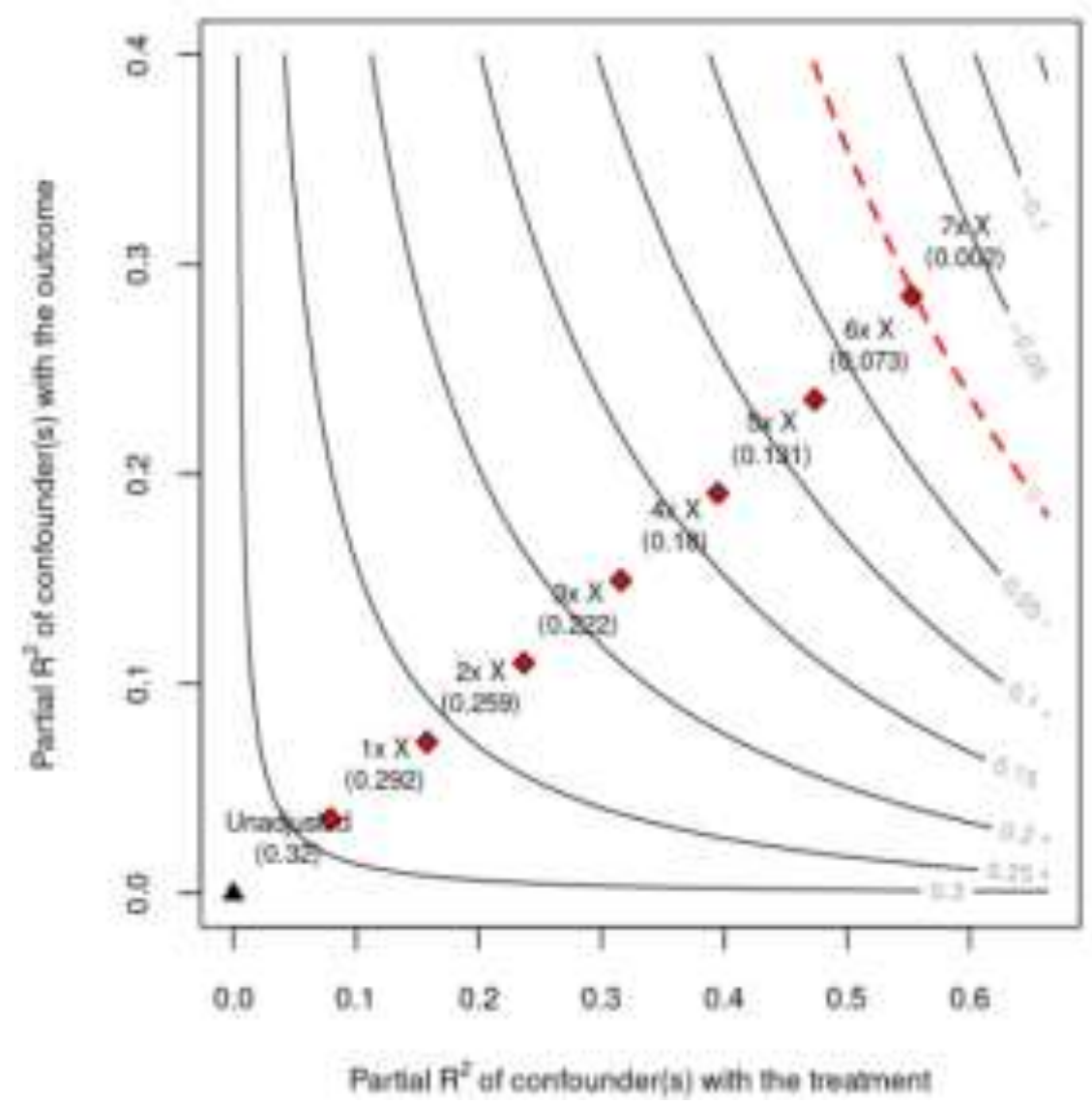

**Figure 5. Contour Plot of Sensitivity Analysis Cinelli and Hazlett (2020)**

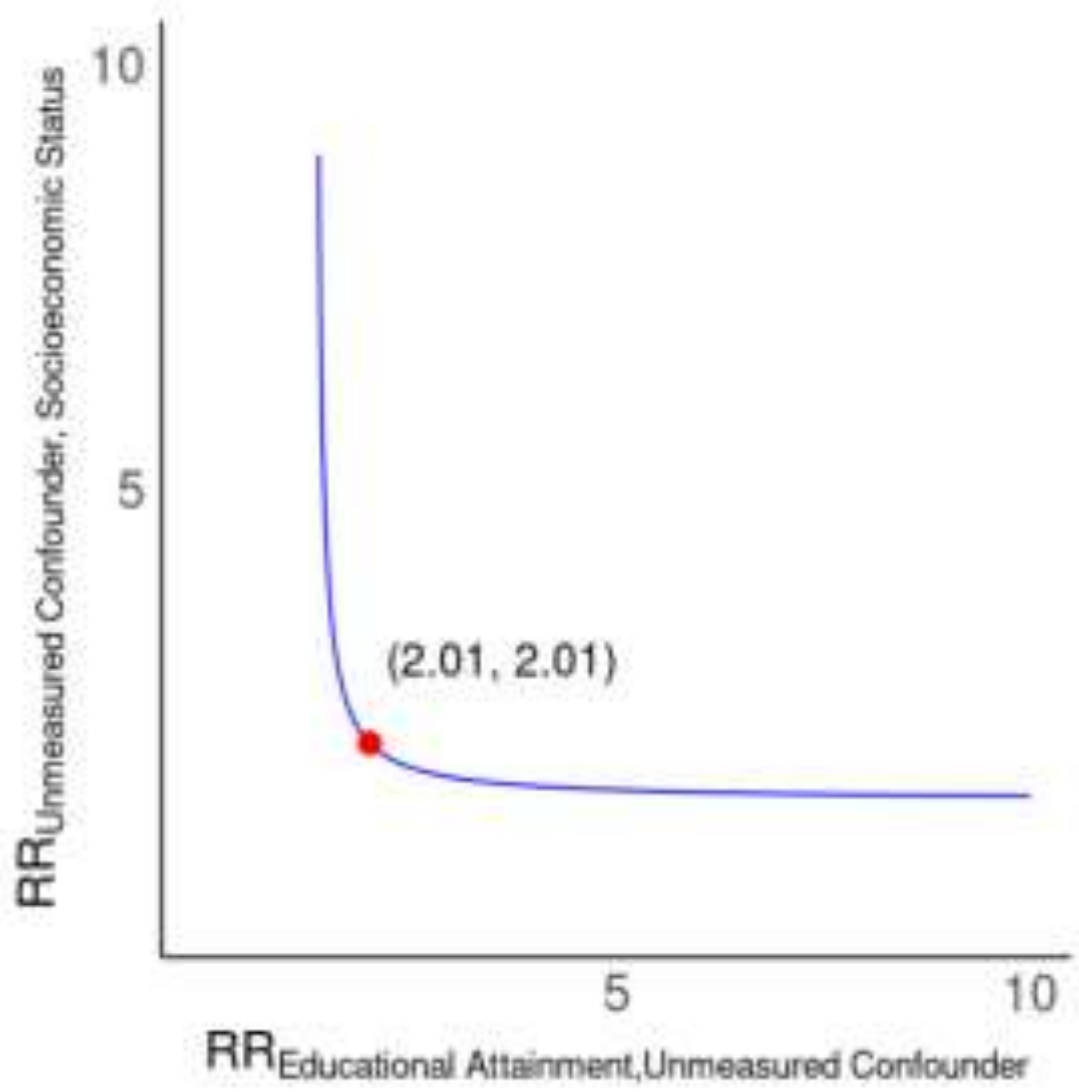

**Figure 6. E-value of the Effect Son's Educational Attainment (T) on Son's Socioeconomic Status in 1962 (Y)**

## Appendices

### A. The query for the search on the prevalence of usage of "sensitivity analysis"

**1.The query for the search in the Abstract, Title, or Keywords fields of a record in the Web of Science (The prevalence of usage of "sensitivity analysis" and its relevant terms in these categories):**

TS= ("sensitivity analysis" OR "bias analysis" OR "omitted variable bias*" OR "coefficient stability" OR "coefficient stability test*" OR "falsification test*" OR "E value*" OR "E-value*") AND WC= ("political science" OR "Economics" OR "sociology" OR "Public, Environmental & Occupational Health") AND PY=(2002-2022)

And the query link is:
https://www.webofscience.com/wos/woscc/summary/7b27f24d-0656-41b3-8aef-e4bd4d18457d-9d49bc54/relevance/1

**2.The query for the search in the Web of Science categories in the Web of Science (all papers in these categories):**

WC= ("political science" OR "Economics" OR "sociology" OR "Public, Environmental & Occupational Health") AND PY=(2002-2022)

And the query link is:
https://www.webofscience.com/wos/woscc/summary/99fb50b2-8715-4493-bf13-7ed9b9ce8eaa-9d4910d2/relevance/1

### B. Review of sociology articles

**Questionnaires**

**1. Does the paper have clear causal claim? (**The amount of causal thinking? ** Observe that this question is not about the amount of causal language but more about the causal thinking. Even if they fail to identify the estimand, but acknowledge it)**

(A). high amount of causal thinking = the study precisely uses at least one of the following elements: Econometric, DAG or potential outcomes to clarify the research (AND uses English to theorize about causal)

(B). moderate amount of causal thinking = uses only states statistical assumption in causal language (e.g., "non-correlated errors" or exogenous vs endogenous variables,), but not clarify causal assumptions (AND tends not to use English to theorize about causal or statistical dependency)

(C). low amount of causal thinking = the study mentions the noun or verb *causality* or *cause* or *mechanisms* or *impact* but no DAG, potential outcomes, or statistical assumptions in causal langue (these are probably most Model-based methods)

(D). no amount of causal thinking = uses none of the stated conditions in the high, moderate or low categories. (these are mostly "Model-based-methods studies")

**2. Does the paper discuss the unmeasured confounders?**

(A) Yes, clearly discussed possible unmeasured confounders. (1)
(B) Mentioned the existence of unmeasured confounders but did not fully take that influence into account empirically (2)
(C) No (3)

**3. Does the paper identify the difference between robustness check (model uncertainty) and sensitivity analysis (confounding uncertainty)?**

(A) Clear difference between the robustness check and sensitivity analysis (1)
(B) Confused the difference or did not identify the difference (2)
(C) Used other terminology but did identify the difference (3)
(D) Used other terminology but did not identify the difference (4)

**4. Does the paper conduct robustness check in terms of model uncertainty?**

(A)Yes (B) No

**5. Does the paper conduct sensitivity analysis in terms of confounding uncertainty?**

(A)Yes (B) No

**6. What is the estimation method?**

**7. Does the paper conduct randomization?**

(A)Yes (B) No

Table 1. Review of sociology articles in AJS and ASR

| | Causal Claim | Discussing unobserved | Differentiating between robustness and sensitivity | Using robustness | Using any sensitivity analysis | Empirical Estimands | Randomized |
|---|---|---|---|---|---|---|---|

| (Drouhot 2021) | 4 | 3 | 2 | YES | NO | Fuzzy logic | NO |
|---|---|---|---|---|---|---|---|
| (Torche and Abufhele 2021) | 2 | 1 | 4 | YES | YES | FE | NO |
| (Campero 2021) | 4 | 2 | 2 | YES | NO | OLS | NO |
| (Duxbury 2021) | 2 | 2 | 4 | YES | NO | KHB | NO |
| (Kalmijn and Leopold 2021) | 4 | 3 | 2 | YES | NO | Event history model | NO |
| (Martínez-Schuldt and Martínez 2021) | 2 | 1 | 4 | YES | NO | multilevel | NO |
| (Leal 2021) | 4 | 3 | 2 | YES | NO | TERG network | NO |
| (Neil and Sampson 2021) | 2 | 1 | 1 | YES | YES | GEE | NO |
| (Schachter, Flores, and Maghbouleh 2021) | 1 | 2 | 2 | NO | NO | experiment | YES |
| (Tilcsik 2021) | 1 | 2 | 2 | NO | NO | experiment | YES |
| (Browning et al. 2021) | 2 | 2 | 4 | YES | NO | FE | NO |
| (Homan and Burdette 2021) | 4 | 1 | 2 | YES | NO | Multiple approaches | NO |
| (McMahan and McFarland 2021) | 2 | 2 | 2 | YES | NO | OLS and network | NO |
| (Chu 2021) | 2 | 1 | 1 | YES | YES | mediation | NO |
| (Zhang and Axinn 2021) | 2 | 3 | 2 | YES | NO | Hazard and FE | NO |
| (Gonalons-Pons and Gangl 2021) | 2 | 1 | 2 | YES | NO | multilevel | NO |
| (Soehl and Karim 2021) | 4 | 3 | 2 | YES | NO | Latent class and logisttic | NO |

| (Torche and Rauf 2021) | 2 | 1 | 2 | YES | NO | FE | NO |
|---|---|---|---|---|---|---|---|
| (Whitham 2021) | 1 | 3 | 4 | NO | NO | mediation | YES |
| (Ashwin, Keenan, and Kozina 2021) | 4 | 1 | 2 | YES | NO | FE | NO |
| (Monk, Esposito, and Lee 2021) | 4 | 3 | 2 | YES | NO | pooled | NO |
| (Gerber and van Landingham 2021) | 4 | 3 | 2 | NO | NO | logistic | NO |
| (Hansen and Toft 2021) | 4 | 3 | 2 | NO | NO | OLS | NO |
| (Maralani and Portier 2021) | 4 | 1 | 2 | YES | NO | logistic | NO |
| (Pfeffer and Waitkus 2021) | 4 | 3 | 2 | YES | NO | Formal decomposition | NO |
| (Bonikowski, Feinstein, and Bock 2021) | 4 | 3 | 2 | YES | NO | logistic | NO |
| (Doering and Ody-Brasier 2021) | 4 | 3 | 4 | YES | NO | FE | NO |
| (Roscigno, Yavorsky, and Quadlin 2021) | 4 | 3 | 2 | YES | NO | logistic | NO |
| (Zhang 2021) | 2 | 2 | 2 | YES | NO | logistic | NO |
| (Bro?i? and Miles 2021) | 2 | 2 | 2 | YES | NO | Random effect | NO |
| (Goldstein and Eaton 2021) | 4 | 2 | 2 | YES | NO | FE pooled | NO |
| (Martin-Caughey 2021) | 4 | 3 | 2 | YES | NO | Linear | NO |
| (Sauer et al. 2021) | 4 | 2 | 2 | YES | NO | LPM | NO |

| (Scarborough et al. 2021) | 4 | 3 | 2 | YES | NO | logistic | NO |
|---|---|---|---|---|---|---|---|
| (Barrie 2021) | 4 | 3 | 2 | YES | NO | FE | NO |
| (Chung, Lee, and Zhu 2021) | 4 | 2 | 2 | YES | NO | FE | NO |
| (Duxbury and Haynie 2021) | 4 | 2 | 2 | YES | NO | FE and dynamic network | NO |
| (Muller and Schrage 2021) | 2 | 2 | 2 | YES | NO | FE | NO |
| (Heiberger, Munoz-Najar Galvez, and McFarland 2021) | 4 | 3 | 2 | YES | NO | Event history | NO |
| (Olzak 2021) | 4 | 2 | 2 | YES | NO | crb | NO |
| (Parolin and Gornick 2021) | 4 | 2 | 2 | YES | NO | decomposition | NO |
| (Passaretta and Skopek 2021) | 2 | 2 | 2 | YES | NO | FE | NO |
| (Wilmers and Aeppli 2021) | 4 | 3 | 2 | YES | NO | FE | NO |

## A Scoping Review of Sociology Studies in AJS and ASR

To evaluate how often sociologists use sensitivity analysis, we collect papers from the ASR and AJS to evaluate the use of sensitivity analysis in terms of the unconfoundedness assumption in sociology, as shown in table 1. Because these two journals are the leading sociological journals, what we find in them provides evidence towards reasoning about unobserved confounding in sociological research. To make the review manageable, our review focuses on Issues 1 to 6 of Volume 86 in ASR and Numbers 4-5 of Volume 126 and Numbers 1-3 of Volume 127 in AJS in 2021.

### 1. Eligibility Criteria

Our review has several selection criteria. Articles are excluded if they pursue qualitative research, theoretical research, methodological research, and computational research without observational data (i.e., studies based on simulation without observational data). Furthermore, if a paper conducted mixed methods that included a quantitative observational part, it will be included. Figure 1 shows the inclusion process of articles. Of the total of 69 articles, 43 papers are quantitative and of which 3 are experimental, mainly survey experiments. Discounting those experimental articles, we have 40 quantitative papers of which 14 are explicitly causal. The next section discusses the differences between these explicitly causal articles and the rest.

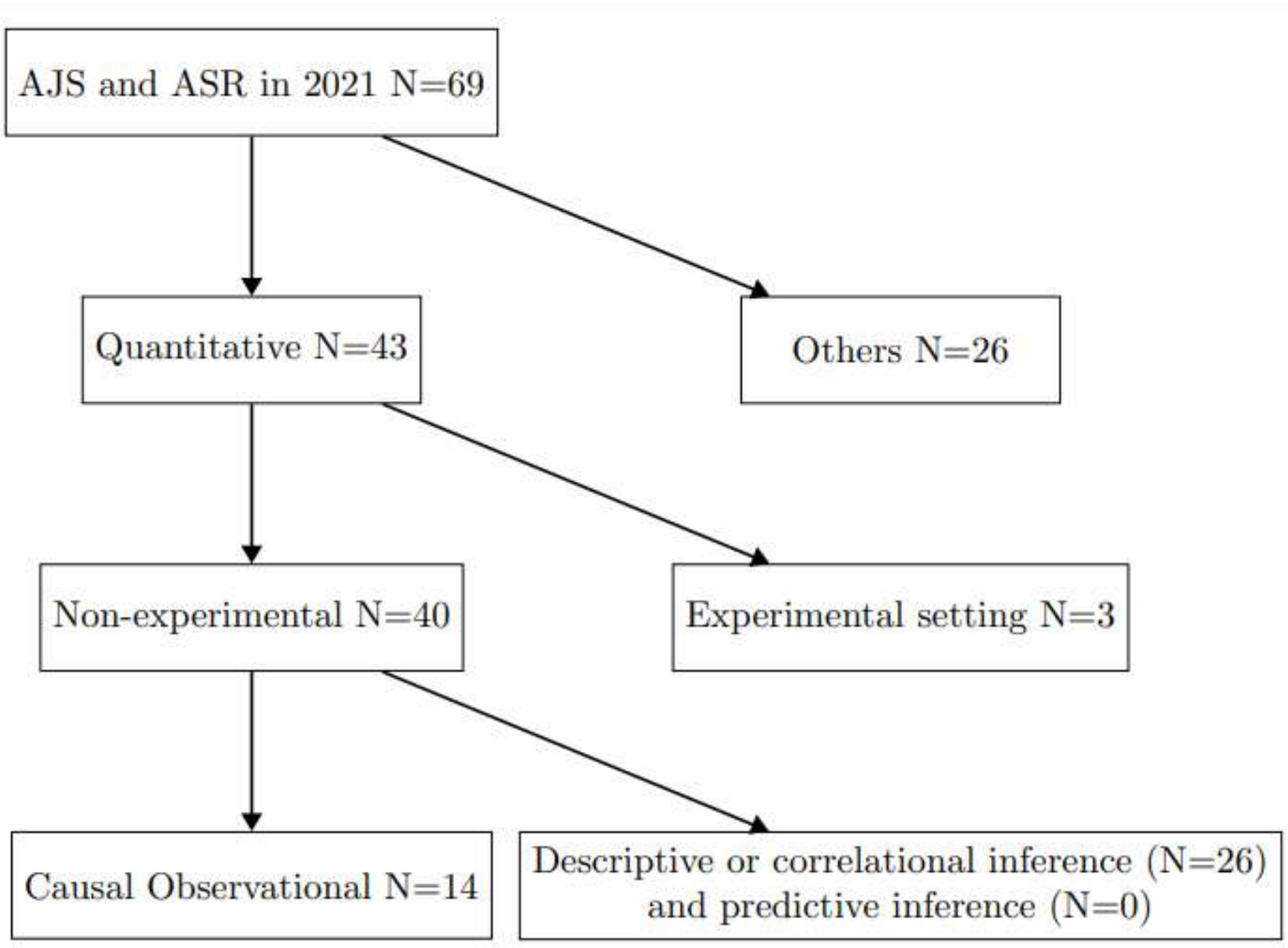

**Figure 1. Flowchart of Review**

2. Precise Causal Quantity

First, we evaluate how precise the causal quantity is described in each selected paper. If a paper discussed the causal unit-specific quantity under a structural causal model or potential outcome framework or followed experimental protocol (Lundberg et al. 2021), it will be categorized into high-level causal quantity type. If a paper includes statistical models and causal assumptions but does not discuss causal unit-specific quantity under these two frameworks precisely, it belongs to the moderate-level causal quantity category. And then, if a paper discusses descriptive unit-specific quantity without clear causal claims or with clear non-causal claims, it will be a low/non-causal quantity paper. The 3 experimental studies with randomization belong to the high-level causal quantity category. And 14 out of 43 studies belong to moderate-level causal quantity because they discussed causal assumptions or statistical models but did not clarify the causal unit-specific quantity precisely under the structural causal model or potential outcome framework. We find that 26 out of 43 studies exhibit low/noncausal quantities because they did not state causal claims and causal assumptions precisely or just provided descriptive unit-specific quantities.

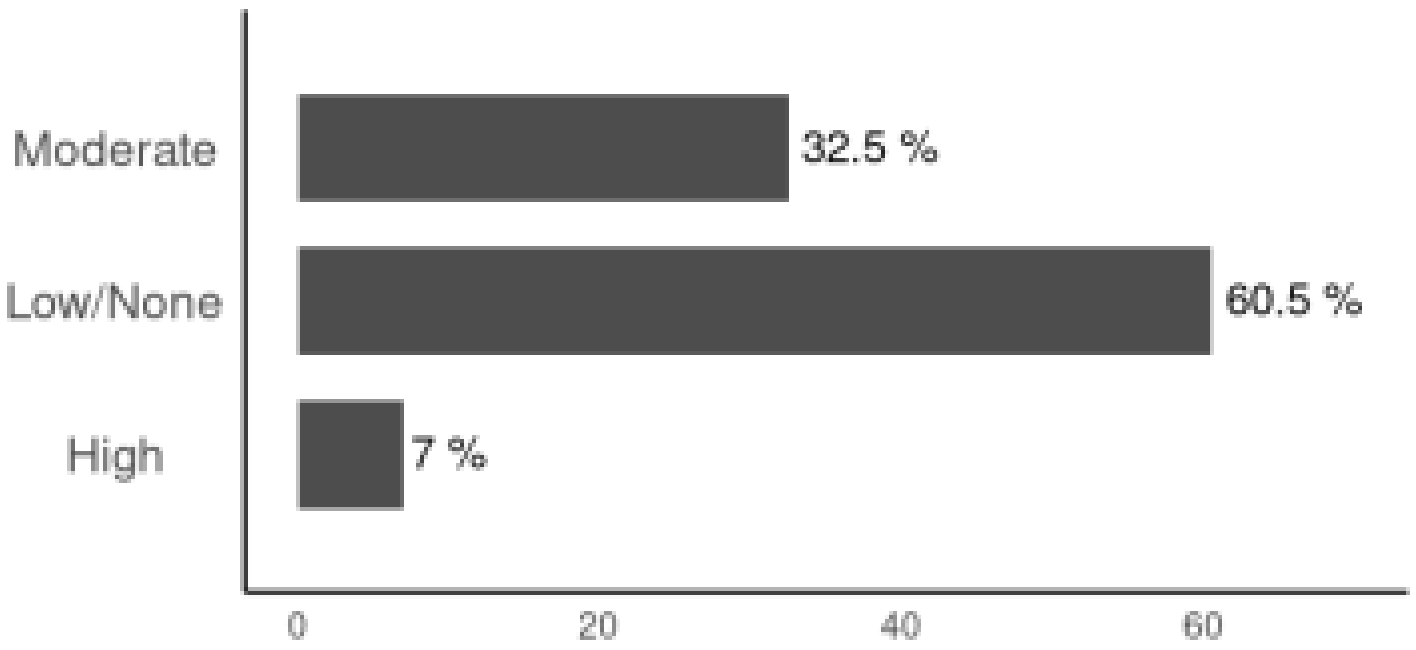

**Figure 2. Evaluation of Precision on Causal Quantity in AJS and ASR**

3. Theorizing about the Unconfoundedness Assumption

Reasoning about the unconfoundedness assumption involves statements if a paper evaluates the existence of unmeasured confounders (or omitted variables) and how the paper evaluates the impacts of unmeasured confounders.

Potential unmeasured confounders or omitted variables that will impact estimates are discussed in 9 out of 40 non-experimental settings. For example, Torche and Abufhele (2021) show how that reasoning unfolds in their study on fertility. They argue that unobserved factors could change nonmarital fertility across municipalities and the possible effects of unmeasured selectivity of individuals 'characteristics and social trends. There are 14 out of 40 papers that mention the existence of unmeasured confounders or omitted variables and were unable to or choose not fully to discuss that influence concretely. The rest of the articles did not discuss unmeasured confounders or omitted variables at all.

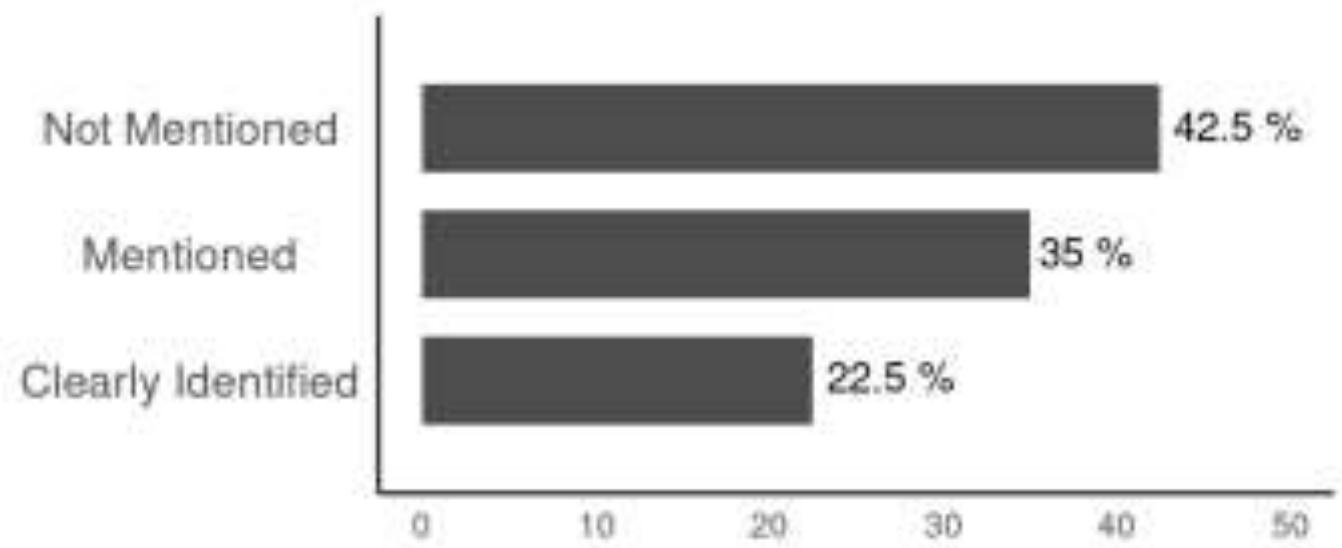

**Figure 3. Evaluation of "Unconfoundedness" or "Omitted Variables" in Practice**

4. The Difference Between Robustness Checks and Sensitivity Analysis

In this step, we evaluated if an article identified the difference in the use of terminology between robustness checks and sensitivity analysis. As previously defined, robustness checks are the evaluation of various model specifications (i.e., different statistical model selections, and different measurements of variables). Figure 4 shows that 2 out of 40 papers identify the differences between robustness checks and sensitivity analysis in the use of terminology precisely. 33 out of 40 papers lacked a clear difference between robustness and sensitivity or did not signal any difference. 5 out of 40 papers used other terminologies to describe robustness checks or sensitivity analysis methods but did not identify the difference. A positive example is Chu (2021) which conducts a mediation analysis with sensitivity analysis to estimate the impacts of unmeasured confounders but also clarified how different model specifications could impact the robustness of estimates.

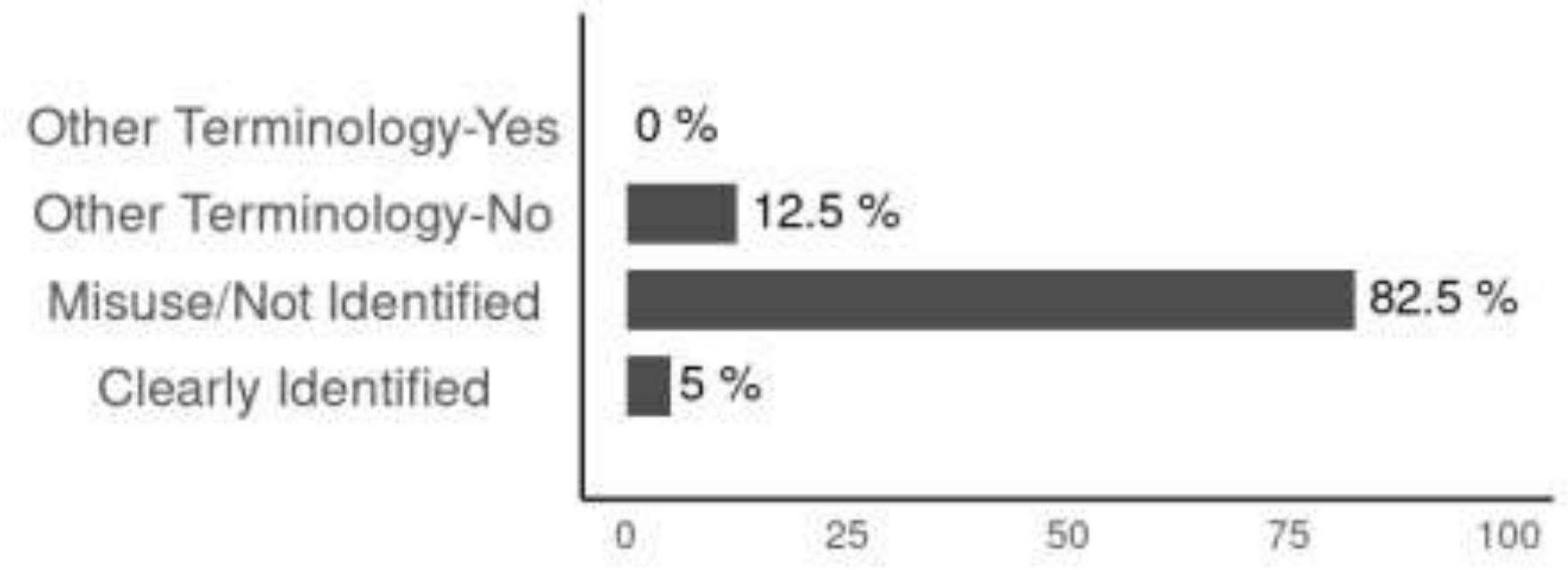

**Figure 4. Evaluation of the Difference Between Robustness Checks and Sensitivity Analysis**

5. The Use of Robustness Checks and Sensitivity Analysis

For the use of robustness checks, 38 out of 40 papers conduct robustness checks to evaluate the effects of various model specifications. Only 2 out of 40 do not clarify the additional robustness checks. These 2 papers test multiple models but did not consider broader model specifications. However, only 3 out of 40 papers conduct a sensitivity analysis, while most papers (92.5%) do not.

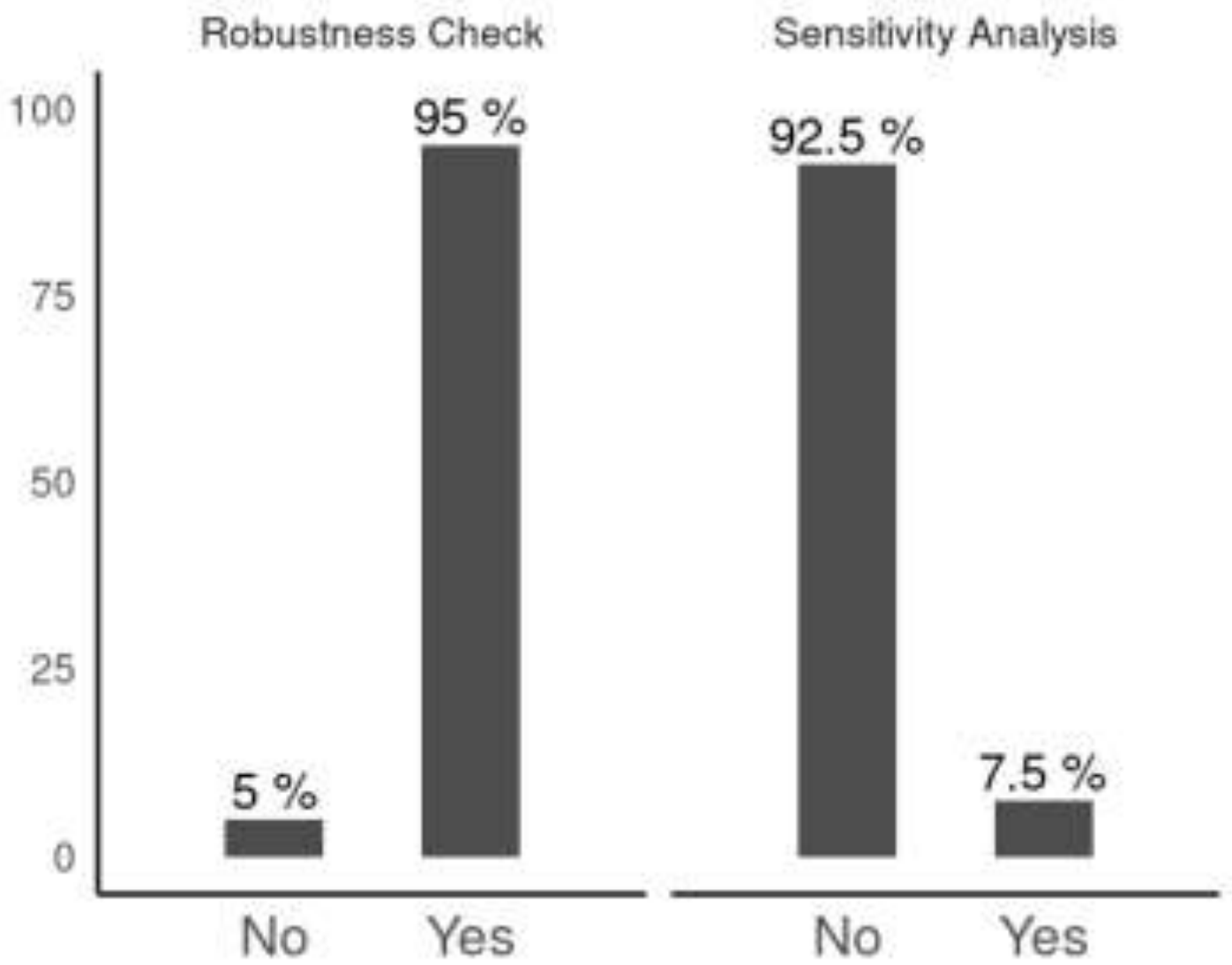

**Figure 5. Evaluation of the Use of Robustness Checks and Sensitivity**

## C. Review of Sensitivity Methods

| Observed confounders | Type of outcome | estimation type |
|---|---|---|
| No | continuous | parametric |
| Yes | Binary | non-parametric |
| No | continuous | Linear parametric |

| | | | alpha and beta indicating the linear structural coefficients of exposure W and Outcome Y respectively. | | | |
|---|---|---|---|---|---|---|
| Mathur, M. B., Ding, P., Riddell, C. A., & VanderWeele, T. J. (2018). Web site and R package for computing E-values. Epidemiology, 29(5), e45-e47. | Pretreatment | Risk Ratio/Risk Difference | The relative risk ratio pair (RREU , RRUD) that indicate the strength of confounding between the exposure E and the outcome D induced by the confounder U. | Yes | Binary | non-parametric |
| Sjölander, A. (2020). A note on a sensitivity analysis for unmeasured confounding, and the related E-value. Journal of Causal Inference, 8(1), 229-248. | Pretreatment | RR/RD | The relative risk ratio pair (RREU , RRUD) that indicate the strength of confounding between the exposure E and the outcome D induced by the confounder U. | Yes | Binary | non-parametric |
| Sjölander, A., & Hössjer, O. (2021). Novel bounds for causal effects based on sensitivity parameters on the risk difference scale. Journal of Causal Inference, 9(1), 190-210. | Pretreatment | Risk Difference= ATE | The relative risk differences (RDEU , RDUD0, RDUD1) that indicate the difference in conditional probabilities of the exposure E and the outcome D induced by the confounder U. | Yes | Binary | non-parametric |
| VanderWeele, T. J., & Arah, O. A. (2011). Bias formulas for sensitivity analysis of unmeasured confounding for general outcomes, treatments, and confounders. Epidemiology, 22(1), 42-52. | Pretreatment | ATE/RD/RR /OR | The relation between U and Y, among those with treatment level A, a1 and a0, within each stratum of X, ie, $\{E(Y \mid a1, x, u)-E(Y \mid a1, x, u)\}$ and $\{E(Y \mid a0, x, u)-E(Y \mid a0, x, u)\}$, as well as how the distribution of the unmeasured confounder U among those with treatment level A=a1 and A=a0 compares | Yes | Binary/categorical/continuous | parametric |

| | | | with the overall distribution of U, within each stratum of X, ie, {P(u\|a1, x) P(u\|x)} and {P(u\|a0, x) P(u\|x)}. | | | |
|---|---|---|---|---|---|---|
| VanderWeele, T. J., & Ding, P. (2017). Sensitivity analysis in observational research: introducing the E-value. Annals of internal medicine, 167(4), 268-274. | Pretreatment | RR/RD | The relative risk ratio pair (RREU , RRUD) that indicate the strength of confounding between the exposure E and the outcome D induced by the confounder U. | Yes | Binary/categorical/continuous | non-parametric |

| Veitch, V., & Zaveri, A. (2020). Sense and sensitivity analysis: Simple post-hoc analysis of bias due to unobserved confounding. Advances in neural information processing systems, 33, 10999-11009. | Pretreatment | ATE | Partial variation in outcome Y and exposure W as a function of two sensitivity parameters alpha and beta indicating the linear structural coefficients of exposure W and Outcome Y respectively. | Yes | continuous | parameteric |
|---|---|---|---|---|---|---|
| VanderWeele, T. J. (2010). Bias formulas for sensitivity analysis for direct and indirect effects. Epidemiology, 21(4), 540-551. | Posttreatment (Mediator and outcome) | CDE/NDE/NIE bias | Difference of conditional expectation of outcome given mediator, confounders (gamma) and difference of the conditional probability of unobserved confounder given meaditor and observed confounder | Yes | Binary/continuous | Linear parametric |
| Imai, K., Keele, L., & Yamamoto, T. (2010). Identification, inference and sensitivity analysis for causal mediation effects. | Pretreatment | NDE/NIE | Correlation Coefficient between the exogenous noise terms of | Yes | continuous | Linear parametric |

| | | | mediator and outcome | | | |
|---|---|---|---|---|---|---|
| Imai, K., Keele, L., & Tingley, D. (2010). A general approach to causal mediation analysis. Psychological methods, 15(4), 309. | Pretreatment | NDE/NIE | Correlation Coefficient between the exogenous noise terms of mediator and outcome | Yes | continuous | Linear parametric |
| Peña, J. M. (2022). Simple yet sharp sensitivity analysis for unmeasured confounding. Journal of Causal Inference, 10(1), 1-17. | Pretreatment | ATE | Maximum and minimum conditional probability of outcome given treatment and the unobserved confounder, for all treatment and the unobserved confounder | No | Binary | non-parametric |
| Franks, A., D'Amour, A., & Feller, A. (2019). Flexible sensitivity analysis for observational studies without observable implications. Journal of the American Statistical Association. | Pretreatment | ATE | sensitivity parameters $\gamma=(\gamma_0,\gamma_1)$, which describe how treatmentassignment depends marginally on each potential outcome | No | Binary, continuous | parametric/ non-parametric |
| Greenland, S. (1996). Basic methods for sensitivity analysis of | Pretreatment | Odds ratio | Prevelence ratios and Odds ratio of | No | Binary | non-parametric |

| biases. International journal of epidemiology, 25(6), 1107-1116. | | | unobserved confounder and treatment and outcome | | | |
|---|---|---|---|---|---|---|
| VanderWeele, T. J., Mukherjee, B., & Chen, J. (2012). Sensitivity analysis for interactions under unmeasured confounding. Statistics in medicine, 31(22), 2552-2564. | Pretreatment | RR/RD Bias | bias formulas to assess the sensitivity of interaction estimates to the presence of an unmeasured confounder. | No | Binary/continuous | non-parametric |
| Cinelli, C., & Hazlett, C. (2020). Making sense of sensitivity: Extending omitted variable bias. Journal of the Royal Statistical Society Series B: Statistical Methodology, 82(1), 39-67. | Pretreatment | ATE bias | Impact of confounder subgroups on outcome (gamma) and Imbalance of confounder with respect to the treatment (delta)<br><br>Reparameterization of gamma and delta into partial R2 variances of unobserved confounder on treatment and outcome in the linear model | Yes | continuous | Linear parametric and its extension to non-linearity |

| Dorie, V., Harada, M., Carnegie, N. B., & Hill, J. (2016). A flexible, interpretable framework for assessing sensitivity to unmeasured confounding. Statistics in medicine, 35(20), 3453-3470. | Pretreatment | ATE | Coefficients of unobserved confounders in the Semi-parameteric model | Yes | continuous | Semi-parametric |
|---|---|---|---|---|---|---|
| Zhang, B., & Tchetgen Tchetgen, E. J. (2022). A semi-parametric approach to model-based sensitivity analysis in observational studies. Journal of the Royal Statistical Society Series A: Statistics in Society, 185(Supplement_2), S668-S691. | Pretreatment | ATE | Coefficients of unobserved confounders in the Semi-parameteric model | Yes | Binary/categorical/continuous | Semi-parametric |
| Balgi, S., Peña, J. M., & Daoud, A. (2022). $\rho $-GNF: A Novel Sensitivity Analysis Approach Under Unobserved Confounders. arXiv preprint arXiv:2209.07111. | Pretreatment | ATE/CATE | Gaussian Copula Correlation Coefficient between the exogenous noise terms of treatment and outcome | Yes | Binary/categorical/continuous | Parametric |
| Cinelli, C., Ferwerda, J., & Hazlett, C. (2020). sensemakr: Sensitivity analysis tools for OLS in R and Stata. Available at SSRN 3588978. | Pretreatment | ATE bias | Impact of confounder subgroups on outcome (gamma) and Imbalance of confounder with respect to the treatment (delta)<br><br>Reparameterization of gamma and delta into partial R2 variances of unobserved confounder on treatment and outcome in the linear model | Yes | continuous | Linear parametric |
| Zheng, J., D'Amour, A., & Franks, A. (2021). Copula-based sensitivity analysis for multi-treatment causal inference with unobserved confounding. arXiv preprint arXiv:2102.09412. | Pretreatment | ATE | The covariance matrix of the Gaussian Copula | Yes | continuous | Linear parametric |
| Zheng, J., D'Amour, A., & Franks, A. (2021). Bayesian inference and partial identification in multi-treatment causal inference | Pretreatment | ATE | Reparameterized structural coefficient gamma with partial R2 | No | continuous | non-parametric |

| with unobserved confounding. arXiv preprint arXiv:2111.07973. | | | outcome variance due to confounder given treatment | | | |
|---|---|---|---|---|---|---|
| Carnegie, N. B., Harada, M., & Hill, J. L. (2016). Assessing sensitivity to unmeasured confounding using a simulated potential confounder. Journal of Research on Educational Effectiveness, 9(3), 395-420. | Pretreatment | ATE/ATT/ATC | Coefficients of unobserved confounders in the Semi-parameteric model | Yes | continuous | Semi-parametric |
| Rosenbaum, P. R., & Rubin, D. B. (1983). Assessing sensitivity to an unobserved binary covariate in an observational study with binary outcome. Journal of the Royal Statistical Society: Series B (Methodological), 45(2), 212-218. | Pretreatment | ATE | Log Odds Ratios of the conditional probabilities of the treatment and outcome given observed confounder | Yes | Binary | parametric |
| Brumback, B. A., Hernán, M. A., Haneuse, S. J., & Robins, J. M. (2004). Sensitivity analyses for unmeasured confounding assuming a marginal structural model for repeated measures. Statistics in medicine, 23(5), 749-767. | Pretreatment/posttreatment | CATE | Unmeasured confounding quantifying function of measured confounders and treatment {c(a,v)} | Yes | continuous | parametric |
| Cochran, W. G., & Rubin, D. B. (1973). Controlling bias in observational studies: A review. Sankhyā: The Indian Journal of Statistics, Series A, 417-446. | Pretreatment | ATE bias | Average value of confounder per each treatment group | No | continuous/discrete | parametric |
| Sjölander, A., & Greenland, S. (2022). Are E-values too optimistic or too pessimistic? Both and neither!. International Journal of Epidemiology, 51(2), 355-363. | Pretreatment | RR/RD | The relative risk ratio pair (RREU , RRUD) that indicate the strength of confounding between the exposure E and the outcome D induced by the confounder U. | Yes | Binary | non-parametric |
| VanderWeele TJ (2013) Unmeasured confounding and hazard scales: sensitivity analysis for total, direct, and indirect effects. Eur J Epidemiol 28(2):113–117. | Posttreatment/ Pretreatment | Hazard Differences and Hazard Ratios | We need to specify the effect of U on the outcome on the hazard difference scale, conditional on the exposure and the covariates; we will call this parameter γ. We also need to specify the difference in the prevalence of U | Yes | survival outcome | parametric |
| | | | amongst the exposed and the unexposed; we will call this parameter δ. | | | |
| Tchetgen Tchetgen EJ, Shpitser I (2012) Semiparametric theory for causal mediation analysis: efficiency bounds, multiple robustness and sensitivity analysis. Ann Stat 40(3):1816–1845 | Pretreatment | NDE/NIE | the proposed approach circumvents this difficulty by concisely encoding a violation of the ignorability assumption for the mediator through the selection bias function tλ (e, m, x), which encodes the magnitude and direction of the unmeasured confounding for the mediator. | Yes | Continuous | semiparametric |

| Lindmark A, de Luna X, Eriksson M (2018) Sensitivity analysis for unobserved confounding of direct and indirect effects using uncertainty intervals. Stat Med 37(10):1744–1762. https://doi.org/10.1002/sim.7620 | Pretreatment | NDE/NIE | correlations between the error terms of the exposure, mediator, and outcome models | No | Binary | parametric |
|---|---|---|---|---|---|---|
| Lindmark A (2019) sensmediation: Parametric estimation and sensitivity analysis of direct and indirect effects. http://cran.R-project.org/package=sensmediation, R package version 0.3.0 | Pretreatment | NDE/NIE | correlations between the error terms of the exposure, mediator, and outcome models | No | Binary | parametric |
| le Cessie S (2016) Bias formulas for estimating direct and indirect effects when unmeasured confounding is present. Epidemiology 27(1):125–132 | Posttreatment (Mediator and outcome) | NDE/NIE bias | Structural coefficients of the linear model and the variance of mediator | Yes | Binary/categorical/continuous | Linear parametric |
| Miao, W., Hu, W., Ogburn, E. L., & Zhou, X. H. (2023). Identifying effects of multiple treatments in the presence of unmeasured confounding. Journal of the American Statistical Association, 118(543), 1953-1967. | Pretreatment | ATE | Leveraging auxiliary variables | Yes | binary/continuous | Linear parametric |

| Kong, D., Yang, S., & Wang, L. (2022). Identifiability of causal effects with multiple causes and a binary outcome. Biometrika, 109(1), 265-272. | Pretreatment | ATE | Leveraging auxiliary variables | Yes | binary/continuous | Exponential parametric |
|---|---|---|---|---|---|---|
| Miao, W., Geng, Z., & Tchetgen Tchetgen, E. J. (2018). Identifying causal effects with proxy variables of an unmeasured confounder. Biometrika, 105 (4), 987-993. | Pretreatment | ATE | proxy variable | Yes | binary/continuous | non-parametric |
| Shi, X., Miao, W., Nelson, J. C., & Tchetgen Tchetgen, E. J. (2020). Multiply robust causal inference with double-negative control adjustment for categorical unmeasured confounding. Journal of the Royal Statistical Society Series B: Statistical Methodology, 82(2), 521-540. | Pretreatment | ATE | Negative control | Yes | binary/continuous | non-parametric |
| Ilse, M., Forré, P., Welling, M., & Mooij, J. M. (2021). Combining interventional and observational data using causal reductions. arXiv preprint arXiv:2103.04786. | Pretreatment | ATE | Joint probability of treatment and outcome for the bounds on interventional distribution - o | Yes | binary/continuous | parametric non-linear |

| Chernozhukov, V., Cinelli, C., Newey, W., Sharma, A., & Syrgkanis, V. (2022). Long story short: Omitted variable bias in causal machine learning (No. w30302). National Bureau of Economic Research. | Pretreatment | ATE bias | Impact of confounder subgroups on outcome (gamma) and Imbalance of confounder with respect to the treatment (delta)<br><br>Reparameterization of gamma and delta into partial R2 variances of unobserved confounder on treatment and outcome in the linear model | Yes | continuous | Non-parametric |
|---|---|---|---|---|---|---|
| Cinelli, C., & Hazlett, C. (2022). An omitted variable bias framework for sensitivity analysis of instrumental variables. Available at SSRN 4217915. | Pretreatment | ATE | Partial R2 (variance) of confounder with outcome and the instrument | Yes | continuous | Linear Parametric |
| Jiang, Y., Kang, H., and Small, D. S. (2018). ivmodel: An r package for inference and sensitivity analysis of instrumental variables | Pretreatment | ATE | delta | Yes | continuous | Linear Parametric |
| models with one endogenous variable |  |  |  |  |  |  |
| Bijlsma, M. J., & Wilson, B. (2020). Modelling the socio-economic determinants of fertility: a mediation analysis using the parametric g-formula. Journal of the Royal Statistical Society Series A: Statistics in Society, 183(2), 493-513. | Posttreatment/Pretreatment | AME | constructing an artificial confounder C, as a function of the variables that it confounds (X and Y) | No | binary/continuous | Linear Parametric |
| Díaz, I., & van der Laan, M. J. (2013). Sensitivity analysis for causal inference under unmeasured confounding and measurement error problems. The international journal of biostatistics, 9(2), 149-160. |  |  |  |  |  |  |
| Amoafo, L., Platz, E., & Scharfstein, D. (2024). Addressing the Influence of Unmeasured Confounding in Observational Studies with Time-to-Event Outcomes: A Semiparametric Sensitivity Analysis Approach. arXiv preprint arXiv:2403.02539. | pretreatment | ATE | A fixed sensitivity analysis parameter that governs deviations from exchangeability. | No | survival outcome | Semiparametric |
| Mameche, S., Vreeken, J., & Kaltenpoth, D. Identifying | pretreatment | ATE | Mutual information | Yes | binary/continuous | nonparametric |

| Confounding from Causal Mechanism Shifts. | | | | | | |
| --- | --- | --- | --- | --- | --- | --- |
| Richardson, A., Hudgens, M. G., Gilbert, P. B., & Fine, J. P. (2014). Nonparametric bounds and sensitivity analysis of treatment effects. Statistical science: a review journal of the Institute of Mathematical Statistics, 29(4), 596. | pretreatment | ATE | Similar to VanderWeele and Arah (2011) | yes | binary/continuous | parametric |
| Yin, M., Shi, C., Wang, Y., & Blei, D. M. (2024). Conformal sensitivity analysis for individual treatment effects. Journal of the American Statistical Association, 119(545), 122-135. | pretreatment | ITE | a way to estimate a range of the ITE under unobserved confounding. The method we develop quantifies unmeasured confounding through a marginal sensitivity model, and adapts the framework of conformal inference to estimate an ITE interval at a given | Yes | binary/continuous | nonparametric |

| | | | confounding strength. | | | |
|---|---|---|---|---|---|---|
| Bonvini, M., & Kennedy, E. H. (2022). Sensitivity analysis via the proportion of unmeasured confounding. Journal of the American Statistical Association, 117(539), 1540-1550. | pretreatment | ATE | the "proportion of unmeasured confounding": the proportion of units for whom the treatment is not as good as randomized even after conditioning on the observed covariates | Yes | binary/continuous | nonparametric |
| Lu, S., & Ding, P. (2023). Flexible sensitivity analysis for causal inference in observational studies subject to unmeasured confounding. arXiv preprint arXiv:2305.17643. | pretreatment | ATE | two SA parameters | Yes | binary/continuous | nonparametric |

## D. Path-specific Effects

In this section, we will redo Blau and Duncan study sensitivity analysis but focus on the path coefficients. Accordingly, in the first and second steps, we define our estimand to be the path-specific effect that produces the outcome: the son's socioeconomic. That entails estimating the effects of the father's socioeconomic status on the son's socioeconomic via the son's educational attainment and the son's socioeconomic status at the first job conditioning on the father's educational attainment. In a linear causal system assuming no interaction effects, the average total effect of the father's socioeconomic status ($X$) on the son's socioeconomic status in 1962 ($Y$) via the son's educational attainment ($T$) and the son's socioeconomic status at the first job ($W$) can be decomposed into three components by following the DAG in Figure 4:

$$Average\ Total\ Effect = \tau_{X\to Y} + \tau_{X\to T\rightsquigarrow Y} + \tau_{X\to W\to Y}$$

Where $\tau_{X\to Y}$ , $\tau_{X\to T\to Y}$ and $\tau_{X\to W\to Y}$ represent the direct effect of father's socioeconomic status on the son's socioeconomic status, the indirect effect via son's educational attainment and the indirect effect via the son's socioeconomic status at the first job, respectively. To simplify this analysis, we convert the father's socioeconomic status from continuous to binary based on its median, indicating the high father's socioeconomic status and the low father's socioeconomic status. Father's educational attainment is the observable confounder. However, the data lacks measurements of the mother's socioeconomic status, their extended family, and friends' status—factors that may likely affect these path-specific effects, which are possible unmeasured confounders.

Figure 7 shows the estimated effects. We use the estimator developed by Zhou and Yamamoto (2022), which assumes the binary unmeasured confounder under the bias formulas developed by VanderWeele (2010). The estimated direct effect of the father's socioeconomic status is about one-third of the total effect, and the estimated combined mediating effect of the son's educational attainment and the son's socioeconomic status at the first job accounts for the remaining effects, but son's educational attainment play a crucial role in combined mediating effects.

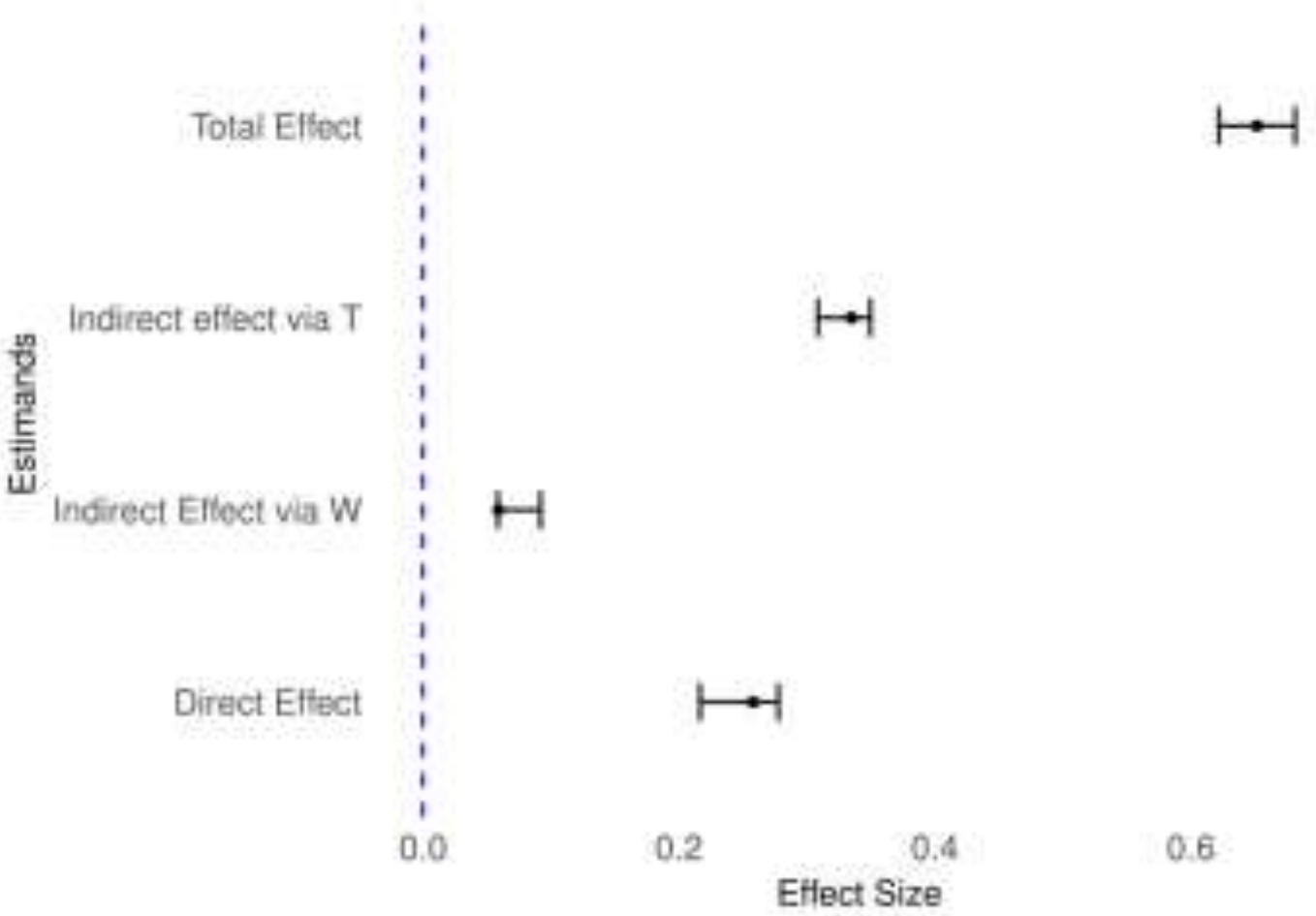

Notes: The point statistical metrics are bootstrapped. Consequently, the bootstrap confidence intervals can be asymmetric if the data is skewed. The total effect of father's socioeconomic status ($X$) on the son's socioeconomic status in 1962 ($Y$) via the son's educational attainment ($T$), the son's socioeconomic status at the first job ($W$) and its direct effect.

**Figure 7. Estimated Effects**

These are all effects that assume away the presence of unmeasured confounding. In the third step, the scales of outcome and mediator in this setting are both continuous. We conduct sensitivity analysis to explore the potential influence of unmeasured confounders between mediators and outcomes, assuming unobserved confounder that affects both the mediators using the bias formula approach (VanderWeele and Arah 2011; Zhou and Yamamoto 2022).

The fourth step is what functional form assumptions one is inclined to make. In this path-analysis setting, current sensitivity methods require the following assumptions with a parametric form: (1) the unmeasured confounder is binary (2) the average "effect" of the unmeasured confounder on the son's socioeconomic status, conditional on father's educational attainment as the baseline covariates, the father's socioeconomic status as the treatment, and two ordered mediators: son's educational attainment and son's socioeconomic status at the first job is constant, which we denote by $\gamma_k$. (3) the difference in the prevalence of the unmeasured confounders between the high and the low father's socioeconomic status, conditional on the father's educational attainment, and two mediators: the son's educational attainment and the son's socioeconomic status at the first job, is constant, which we denote by $\eta_k$ (VanderWeele and Arah 2011; Zhou and Yamamoto 2022).

In the fifth step, we interpret how strong unmeasured confounding must be to overthrow our result. Figure 8 shows the contour plot for sensitivity analysis for the first mediator: the son's educational attainment (0.34, 95%CI: 0.31,0.36), as previously suggested, one possible unmeasured confounder could be genetic propensity that affects the son's educational attainment and son's socioeconomic status in 1962. The contours represent the bias-adjusted estimates of the PSE via the son's educational attainment, plotted as a function of $\gamma_1$ and $\eta_1$. The parameter $\gamma_1$ denotes the effect of unmeasured confounder on the son's socioeconomic status in 1962, conditional on the father's socioeconomic status, the son's educational attainment, and the father's educational attainment, and $\eta_1$denotes the difference in the prevalence of the unmeasured confounder between units with high and low father's socioeconomic status conditional on the son's educational attainment, and the father's educational attainment. For instance, Figure 8 shows that the contour plot shifts its appearance to indicate the estimate shifts sign when $\eta_1 = 0.3$ and $\gamma_1 = -1$. In other words, if the difference in son's socioeconomic status in 1962 compard with those with unmeasured confounder $U = 1$ and $U = 0$ is -1 across the treatment groups, and the difference in the prevalence of the unmeasured confounder between units with high and low father's socioeconomic status is 0.3, the estimated effects of the son's educational attainment will shift its sign.

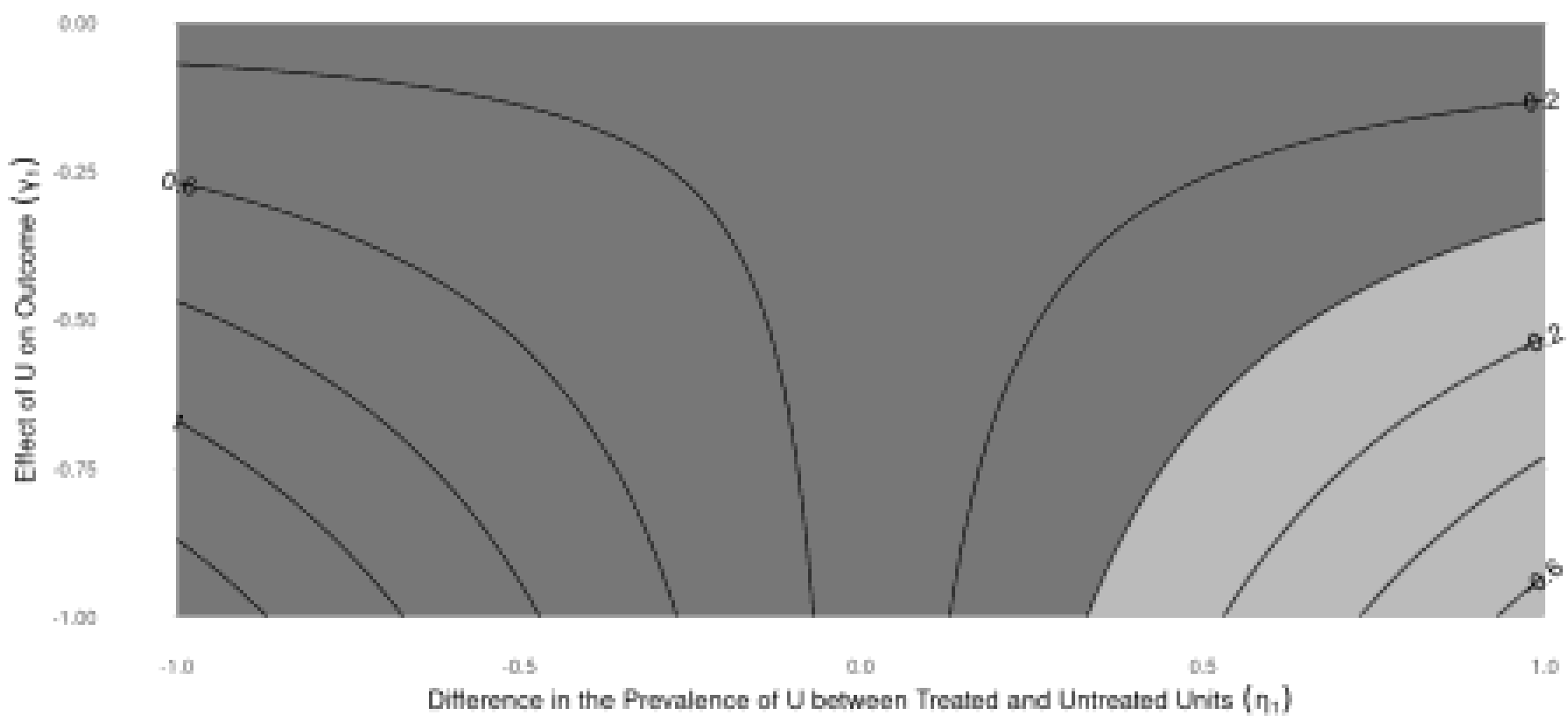

**Figure 8. Bias-adjusted Estimates of the Path-specific Effect of the Father's Socioeconomic Status on the Son's Socioeconomic Status in 1962 via the Son's Educational Attainment**

## E. Codes for Empirical Illustrations in Section 4

```
setwd("data source")
### Read data
#U  # Son's Educational Attainment
#V  # Father's Educational Attainment
#W  # Son's SocioEconomicStatus 1st Job
#X  # Father's SocioEconomicStatus
#Y  # Son's SocioEconomicStatus 1962

duncan_data = read.csv("blau_duncan_1967_cgnf.csv")
duncan_data$X2 = ifelse(duncan_data$X > median(duncan_data$X), 1, 0)
duncan_data$X = scale(duncan_data$X)
duncan_data$Y = scale(duncan_data$Y)
### Model estimation for ATE (4.1 case)
model1 = lm(Y ~ U + V + X, data = duncan_data)
```

```r
### Check the results of estimation
summary(model1)

### Omitted Variable Bias
library("sensemakr")
## Carlos Cinelli and Chad Hazlett (2020). Making Sense of Sensitivity: Extending O
mitted Variable Bias. Journal of the Royal Statistical Society, Series B (Statistical Met
hodology).

darfur.sensitivity <- sensemakr(
  model = model1,
  treatment = "U",
  benchmark_covariates = "X",
  kd = 1:7,
  ky = 1:7,
  q = 1,
  alpha = 0.05,
  reduce = TRUE
)
ovb_minimal_reporting(darfur.sensitivity, format = "latex")
### Plot to get Figure 5
plot(darfur.sensitivity)

### Evalue
library("EValue")
```

```r
est = OLS(model1$coefficients[2], sd = sd(duncan_data$Y))
# for a 1-unit increase in treatment
evalue(
  est = est,
  se = summary(model1)$coefficients['U', 'Std. Error'],
  sd = summary(model1)$sigma
)

##             point    lower    upper
## RR       1.339605 1.329041 1.350253
## E-values 2.014095 1.990334       NA

RR = 1.34
xmax = 10
library(ggplot2)

# Adjust RR if it's less than 1 (as per your condition)
if (RR < 1) {
  RR <- 1 / RR
}

# Preparing data for the plot
x_main <- seq(RR, xmax, by = 0.01)
y_main <- RR * (RR - 1) / (x_main - RR) + RR
```

```r
data_main <- data.frame(x = x_main, y = y_main)

# Calculating the special 'high' point
high <- RR + sqrt(RR * (RR - 1))
high_point_data <- data.frame(x = high, y = high)

# Creating the plot
p <- ggplot() +
  # Adding the main line
  geom_line(data = data_main, aes(x = x, y = y), color = "blue") +
  # Highlighting the 'high' point
  geom_point(data = high_point_data,
             aes(x = x, y = y),
             color = "red",
             size = 3) +
  # Annotating the 'high' point
  geom_text(
    data = high_point_data,
    aes(x = x + 3, y = y + 1),
    label = sprintf("(%.2f, %.2f)", high, high),
    hjust = 1,
    size = 5
  ) +
  # Customizing the scales and labels
  scale_x_continuous(breaks = seq(5, xmax, by = 5), limits = c(0, xmax)) +
  scale_y_continuous(breaks = seq(5, xmax, by = 5), limits = c(0, xmax)) +
  labs(x = bquote(RR["Educational Attainment,Unmeasured Confounder"]),
       y = bquote(RR["Unmeasured Confounder, Socioeconomic Status"]),
       title = "") +
  # Applying theme and customizations
  theme_minimal() +
  theme(
    axis.title = element_text(size = 16),
    axis.text = element_text(size = 16),
    plot.title = element_text(size = 14, face = "bold"),
    axis.line = element_line(color = "black"),
    panel.grid.major = element_blank(),
    panel.grid.minor = element_blank(),
    legend.position = "none"
  ) +
  coord_fixed(ratio = 1)
# Display the plot for Figure 6
print(p)
```

**E.Toy Model for ATE with/without the Unmeasured Confounder**

```r
# Set seed for reproducibility
set.seed(1234)

# Parameters for the true model
```

```r
beta_0 <- 2
beta_T <- 3 #true ATE
beta_U <- 4

# Generate binary treatment (T) and unmeasured confounder (U)
n <- 1000  # Sample size
T <- rbinom(n, 1, 0.5)  # Random treatment assignment
U <- rbinom(n, 1, 0.5)  # Random unmeasured confounder

# Generate outcome variable based on the true model
epsilon <- rnorm(n)  # Random noise
Y <- beta_0 + beta_T * T + beta_U * U + epsilon

# Create data frame
data <- data.frame(T, U, Y)

# Calculate the true expectations manually, conditioning on U
Y10 <- mean(data$Y[data$T == 1 & data$U == 0])
Y11 <- mean(data$Y[data$T == 1 & data$U == 1])
Y00 <- mean(data$Y[data$T == 0 & data$U == 0])
Y01 <- mean(data$Y[data$T == 0 & data$U == 1])

# Calculate the probabilities of U
PU0 <- mean(data$U == 0)
PU1 <- mean(data$U == 1)

# Calculate the true ATE by hands
true_ate <- (Y10 * PU0 + Y11 * PU1) - (Y00 * PU0 + Y01 * PU1)

# Manually calculate the expected values of U given T
U1 <- mean(data$U[data$T == 1])
U0 <- mean(data$U[data$T == 0])

# Calculate the naive ATE without accounting for U
naive_T1 <- mean(data$Y[data$T == 1])
naive_T0 <- mean(data$Y[data$T == 0])
naive_ate <- naive_T1 - naive_T0

# Calculate the bias
bias <- beta_U * (U1 - U0)

# Estimate ATE without accounting for the unmeasured confounder
lm_naive_model <- lm(Y ~ T, data = data)
lm_naive_ate <- coef(lm_naive_model)["T"]

# Estimate ATE accounting for the unmeasured confounder
lm_adjusted_model <- lm(Y ~ T + U, data = data)
lm_adjusted_ate <- coef(lm_adjusted_model)["T"]
```

```
# Print results
cat("ATE (the setting):",beta_T, "\n")

## ATE (the setting): 3

cat("ATE (calculated manually, conditioning on U):", true_ate, "\n")

## ATE (calculated manually, conditioning on U): 3.046925

cat("Naive ATE (without unmeasured confounder):", naive_ate, "\n")

## Naive ATE (without unmeasured confounder): 3.235395

cat("ATE (OLS, conditioning on U):", lm_adjusted_ate, "\n")

## ATE (OLS, conditioning on U): 3.047023

cat("Naive ATE (OLS,without unmeasured confounder):", lm_naive_ate, "\n")

## Naive ATE (OLS,without unmeasured confounder): 3.235395

cat("Bias due to unmeasured confounder:", bias, "\n")

## Bias due to unmeasured confounder: 0.1885644
```